%% file: main.tex
\documentclass[trackchanges,twocolumn]{aastex701}

\usepackage{amsmath,amsfonts,amssymb}
\usepackage{graphicx}
\usepackage{natbib}
\usepackage{enumitem}
\usepackage{hyperref}
\input{commands}
\usepackage{fontspec}

\usepackage[english]{babel} 
\babelprovide{punjabi} 
\babelfont[punjabi]{rm}[
  Renderer=Harfbuzz,
  Script=Gurmukhi
]{FreeSerif.otf} 

\newcommand{\Punj}[1]{\selectlanguage{punjabi}{#1}\selectlanguage{english}}

\begin{document}

\title{THRILS -- The High-(Redshift+Ionization) Line Search: \\ Program Description \& Redshift Catalog}


\suppressAffiliations


\author[0000-0001-6251-4988]{Taylor A.\ Hutchison}
\altaffiliation{NASA Postdoctoral Fellow}
\affiliation{Authors contributed equally to this work.}
\affiliation{Astrophysics Science Division, Code 660, NASA Goddard Space Flight Center, 8800 Greenbelt Rd., Greenbelt, MD 20771, USA}
\email[show]{taylor.hutchison@nasa.gov}

\author[0000-0003-2366-8858]{Rebecca L.\ Larson}
\altaffiliation{Giacconi Postdoctoral Fellow}
\affiliation{Authors contributed equally to this work.}
\affil{Space Telescope Science Institute, 3700 San Martin Drive, Baltimore, MD 21218, USA}
\email[show]{rlarson@stsci.edu}


\author[0000-0002-7959-8783]{Pablo Arrabal Haro}
\altaffiliation{NASA Postdoctoral Fellow}
\affiliation{Astrophysics Science Division, Code 660, NASA Goddard Space Flight Center, 8800 Greenbelt Rd., Greenbelt, MD 20771, USA}
\email{pablo.arrabalharo@nasa.gov}

\author[0000-0003-3216-7190]{Erini Lambrides}
\altaffiliation{NASA Postdoctoral Fellow}
\affiliation{Astrophysics Science Division, Code 660, NASA Goddard Space Flight Center, 8800 Greenbelt Rd., Greenbelt, MD 20771, USA}
\email{erini.lambrides@nasa.gov}

\author[0000-0003-4922-0613]{Katherine Chworowsky}
\affiliation{Department of Astronomy, The University of Texas at Austin, Austin, TX 78712}
\affiliation{Cosmic Frontier Center, The University of Texas at Austin, Austin, TX 78712}
\email{k.chworowsky@utexas.edu}

\author[0000-0002-3475-7648]{Gourav Khullar (\Punj{ਗੌਰਵ ਖੁੱਲਰ})}
\affiliation{Department of Astronomy, University of Washington, Physics-Astronomy Building, Box 351580, Seattle, WA 98195-1700, USA}
\affiliation{Institute for Data-Intensive Research in Astrophysics and Cosmology (DiRAC), University of Washington, Physics-Astronomy Building, Box 351580, Seattle, WA 98195-1700, USA}
\affiliation{eScience Institute, University of Washington, Physics-Astronomy Building, Box 351580, Seattle, WA 98195-1700, USA}
\email{gkhullar@uw.edu}

\author[0000-0001-8047-8351]{Kelcey Davis}
\altaffiliation{NSF Graduate Research Fellow}
\affiliation{Department of Physics, 196A Auditorium Road, Unit 3046, University of Connecticut, Storrs, CT 06269, USA}
\email{kelcey.davis33@gmail.com}

\author[0000-0001-8519-1130]{Steven L.\ Finkelstein}
\affiliation{Department of Astronomy, The University of Texas at Austin, Austin, TX 78712}
\affiliation{Cosmic Frontier Center, The University of Texas at Austin, Austin, TX 78712}
\email{stevenf@astro.as.utexas.edu}

\author[0000-0002-7627-6551]{Jane R.\ Rigby}
\affiliation{Astrophysics Science Division, Code 660, NASA Goddard Space Flight Center, 8800 Greenbelt Rd., Greenbelt, MD 20771, USA}
\email{jane.r.rigby@nasa.gov}

\author[0000-0001-6813-875X]{Guillermo Barro}
\affiliation{University of the Pacific, Stockton, CA 90340 USA}
\email{gbarro@pacific.edu}

\author[0000-0001-7151-009X]{Nikko J.\ Cleri}
\affiliation{Department of Astronomy and Astrophysics, The Pennsylvania State University, University Park, PA 16802, USA}
\affiliation{Institute for Computational \& Data Sciences, The Pennsylvania State University, University Park, PA 16802, USA}
\affiliation{Institute for Gravitation and the Cosmos, The Pennsylvania State University, University Park, PA 16802, USA}
\email{cleri@psu.edu}

\author[0000-0002-8360-3880]{Dale Kocevski}
\affiliation{Department of Physics and Astronomy, Colby College, Waterville, ME 04901, USA}
\email{dale.kocevski@colby.edu}


\author[0000-0002-0243-6575]{Jacqueline Antwi-Danso}
\altaffiliation{Dunlap Fellow}
\affiliation{Department of Astronomy \& Astrophysics, University of Toronto, Toronto, Ontario M5S 3H4, Canada}
\email{j.antwidanso@utoronto.ca}

\author[0000-0002-9921-9218]{Mic Bagley}
\affiliation{Astrophysics Science Division, Code 660, NASA Goddard Space Flight Center, 8800 Greenbelt Rd., Greenbelt, MD 20771, USA}
\email{micaela.bagley@nasa.gov}

\author[0000-0002-4153-053X]{Danielle A.\ Berg}
\affiliation{Department of Astronomy, The University of Texas at Austin, Austin, TX 78712}
\affiliation{Cosmic Frontier Center, The University of Texas at Austin, Austin, TX 78712}
\email{daberg@austin.utexas.edu}

\author[0000-0003-0212-2979]{Volker Bromm}
\affiliation{Department of Astronomy, The University of Texas at Austin, Austin, TX 78712}
\affiliation{Cosmic Frontier Center, The University of Texas at Austin, Austin, TX 78712}
\email{vbromm@astro.as.utexas.edu}

\author[0000-0003-2332-5505]{Oscar Chavez Ortiz}
\affiliation{Department of Astronomy, The University of Texas at Austin, Austin, TX 78712}
\affiliation{Cosmic Frontier Center, The University of Texas at Austin, Austin, TX 78712}
\email{chavezoscar009@utexas.edu}

\author[0000-0002-0302-2577]{John Chisholm}
\affiliation{Department of Astronomy, The University of Texas at Austin, Austin, TX 78712}
\affiliation{Cosmic Frontier Center, The University of Texas at Austin, Austin, TX 78712}
\email{chisholm@austin.utexas.edu}

\author[0000-0003-3038-8045]{Sadie C.\ Coffin}
\affiliation{Laboratory for Multiwavelength Astrophysics, School of Physics and Astronomy, Rochester Institute of Technology, 84 Lomb Memorial Drive, Rochester, NY 14623, USA}
\email{scc3577@rit.edu}

\author[0000-0003-1371-6019]{M. C. Cooper}
\affiliation{Department of Physics \& Astronomy, University of California, Irvine, 4129 Reines Hall, Irvine, CA 92697, USA}
\email{cooper@uci.edu}

\author[0000-0003-3881-1397]{Olivia Cooper}
\altaffiliation{NSF Postdoctoral Research Fellow}
\affiliation{Department for Astrophysical \& Planetary Science, University of Colorado, Boulder, CO 80309, USA}
\email{olivia.cooper@colorado.edu}

\author[0000-0002-1803-794X]{Isa G.\ Cox}
\affiliation{Laboratory for Multiwavelength Astrophysics, School of Physics and Astronomy, Rochester Institute of Technology, 84 Lomb Memorial Drive, Rochester, NY 14623, USA}
\email{igc5972@rit.edu}

\author[0000-0001-5414-5131]{Mark Dickinson}
\affiliation{NSF NOIRLab, 950 N.\ Cherry Ave., Tucson, AZ 85719, USA}
\email{mark.dickinson@noirlab.edu}

\author[0000-0001-7113-2738]{Harry Ferguson}
\affil{Space Telescope Science Institute, 3700 San Martin Drive, Baltimore, MD 21218, USA}
\email{ferguson@stsci.edu}

\author[0000-0002-3560-8599]{Maximilien Franco}
\affiliation{CEA, Universit\'e Paris-Saclay, Universit\'e Paris Cit\'e, CNRS, AIM, 91191, Gif-sur-Yvette, France}
\email{maximilien.franco@austin.utexas.edu}

\author[0000-0003-2098-9568]{Jonathan P. Gardner}
\affiliation{Sciences and Exploration Directorate, NASA Goddard Space Flight Center, 8800 Greenbelt Rd, Greenbelt, MD 20771, USA}
\email{jonathan.p.gardner@nasa.gov}

\author[0009-0006-1252-206X]{Ananya Ganapathy}
\affiliation{Department of Physics \& Astronomy, Johns Hopkins University, Baltimore, MD 21218, USA}
\email{aganapa1@jh.edu}

\author[0000-0001-9440-8872]{Norman Grogin}
\affil{Space Telescope Science Institute, 3700 San Martin Drive, Baltimore, MD 21218, USA}
\email{nagrogin@stsci.edu}

\author[0000-0002-3301-3321]{Michaela Hirschmann}
\affiliation{Institute of Physics, Lab for galaxy evolution, EPFL, Observatoire de Sauverny, Chemin Pegasi 51, 1290 Versoix, Switzerland}
\email{michaela.hirschmann@epfl.ch}

\author[0000-0002-1416-8483]{Marc Huertas-Company}
\affiliation{Instituto de Astrof\'ısica de Canarias (IAC), La Laguna, E-38205, Spain}
\affiliation{Universidad de La Laguna. Avda. Astrof\'ısico Fco. Sanchez, La La- guna, Tenerife, Spain}
\affiliation{Observatoire de Paris, LERMA, PSL University, 61 avenue de l’Observatoire, F-75014 Paris, France}
\affiliation{Universit\'e Paris-Cit\'e, 5 Rue Thomas Mann, 75014 Paris, France}
\email{mhuertas@iac.es}

\author[0000-0003-1187-4240]{Intae Jung}
\affil{Space Telescope Science Institute, 3700 San Martin Drive, Baltimore, MD 21218, USA}
\email{ijung@stsci.edu}

\author[0000-0001-9187-3605]{Jeyhan S.\ Kartaltepe}
\affiliation{Laboratory for Multiwavelength Astrophysics, School of Physics and Astronomy, Rochester Institute of Technology, 84 Lomb Memorial Drive, Rochester, NY 14623, USA}
\email{jeyan@astro.rit.edu}

\author[0000-0002-6610-2048]{Anton M.\ Koekemoer}
\affiliation{Space Telescope Science Institute, 3700 San Martin Drive, Baltimore, MD 21218, USA}
\email{koekemoer@stsci.edu}

\author[0000-0003-1581-7825]{Ray A.\ Lucas}
\affil{Space Telescope Science Institute, 3700 San Martin Drive, Baltimore, MD 21218, USA}
\email{lucas@stsci.edu}

\author[0000-0001-8688-2443]{Elizabeth McGrath}
\affiliation{Department of Physics and Astronomy, Colby College, Waterville, ME 04901, USA}
\email{emcgrath@colby.edu}

\author[0000-0003-4965-0402]{Alexa M.\ Morales}\altaffiliation{NSF Graduate Research Fellow}
\affiliation{Department of Astronomy, The University of Texas at Austin, Austin, TX 78712}
\affiliation{Cosmic Frontier Center, The University of Texas at Austin, Austin, TX 78712}
\email{alexa.morales@utexas.edu}

\author[0000-0002-4606-4240]{Grace M.\ Olivier}
\affiliation{The Observatories of the Carnegie Institution for Science, 813 Santa Barbara Street, Pasadena, CA 91101, USA}
\email{golivier@carnegiescience.edu}

\author[0000-0001-7503-8482]{Casey Papovich}
\affiliation{Department of Physics and Astronomy, Texas A\&M University, College Station, TX, 77843-4242 USA}
\affiliation{George P. and Cynthia Woods Mitchell Institute for Fundamental Physics and Astronomy,\\ Texas A\&M University, College Station, TX, 77843-4242 USA}
\email{papovich@tamu.edu}

\author[0000-0000-0000-0000]{Pablo G. P\'erez-Gonz\'alez}
\affiliation{Centro de Astrobiolog\'{\i}a (CAB), CSIC-INTA, Ctra. de Ajalvir km 4, Torrej\'on de Ardoz, E-28850, Madrid, Spain}
\email{pgperez@cab.inta-csic.es}

\author[0000-0003-3382-5941]{Nor Pirzkal}
\affil{Space Telescope Science Institute, 3700 San Martin Drive, Baltimore, MD 21218, USA}
\email{npirzkal@stsci.edu}

\author[0000-0002-6748-6821]{Rachel S. Somerville}
\affiliation{Center for Computational Astrophysics, Flatiron Institute, 162 5th Ave, New York, NY 10010, USA}
\email{rsomerville@flatironinstitute.org}

\author[0000-0003-1282-7454]{Anthony J.\ Taylor}
\affiliation{Department of Astronomy, The University of Texas at Austin, Austin, TX 78712}
\affiliation{Cosmic Frontier Center, The University of Texas at Austin, Austin, TX 78712}
\email{anthony.taylor@austin.utexas.edu}

\author[0000-0002-1410-0470]{Jonathan R.\ Trump}
\affil{Department of Physics, 196A Auditorium Road, Unit 3046, University of Connecticut, Storrs, CT 06269, USA}
\email{jonathan.trump@uconn.edu}

\author[0000-0002-8163-0172]{Brittany Vanderhoof}
\affil{Space Telescope Science Institute, 3700 San Martin Drive, Baltimore, MD 21218, USA}
\email{bvanderhoof@stsci.edu}

\author[0000-0001-6065-7483]{Benjamin Weiner}
\affiliation{MMT/Steward Observatory, University of Arizona, 933 N. Cherry Ave., Tucson, AZ 85721, USA}
\email{bjw@as.arizona.edu}

\author[0000-0003-1815-0114]{Brian Welch}
\affiliation{International Space Science Institute, Hallerstrasse 6, 3012 Bern, Switzerland}
\email{brian.welch@issibern.ch} 

\author[0000-0003-3466-035X]{{L.~Y.~Aaron} {Yung}}
\altaffiliation{Giacconi Postdoctoral Fellow}
\affil{Space Telescope Science Institute, 3700 San Martin Drive, Baltimore, MD 21218, USA}
\email{yung@stsci.edu}

\author[0000-0002-7051-1100]{Jorge A. Zavala}
\affiliation{University of Massachusetts Amherst, 
710 North Pleasant Street, Amherst, MA 01003-9305, USA}
\email{jorgea.zavalas@gmail.com}

\collaboration{all}{the THRILS collaboration}

\shorttitle{THRILS: Program Description \& Redshift Catalog}
\shortauthors{Hutchison \& Larson et al.}

\begin{abstract}

To date, many spectroscopic confirmations of $z>7$ galaxies have been obtained using \jwst/NIRSpec prism observations, with most of their physical properties inferred from these observations and corresponding imaging. What is needed are higher-resolution spectra at deeper depths to study these sources in detail.
We present The High-(Redshift+Ionization) Line Search (THRILS) program: deep ($>$8 hr) observations in two pointings of \jwst/NIRSpec G395M spectroscopy to 1) probe high ionization spectral features in $z>8$ galaxies that are indicative of top-heavy initial mass functions or growing massive black holes, 2) search for accreting supermassive black holes in typical galaxies at $z \sim 4-9$ through broad Balmer line emission, and 3) probe the stellar-mass growth histories of massive galaxies. We include spectroscopic redshift measurements for 89 sources from the THRILS data, as well as a detection threshold for the full and $\sim$half depth integration times of the program.

\end{abstract}


\section{Introduction}\label{sec:intro}

Over the past few years, galaxy evolution has witnessed an explosion of exciting science from \jwst\ \citep{Gardner.2023JWST}, with many discoveries shedding light on previously unreachable observations of early epochs. In the myriad discoveries from the first few years of \jwst's mission, three results stand out for their unexpected nature and their potential to illuminate the earliest phases of galaxy formation \citep[e.g.,][]{bromm2011}.

Initial results from early \jwst\ imaging detected the presence of galaxies at $z >$ 9 \citep[e.g.,][]{adams22, finkelstein22c, atek23}.  While finding such galaxies was one of the primary science goals of \jwst, the high abundance and the existence of very bright sources out to photometric redshifts of 17 were unexpected \citep{donnan22,harikane22}. While the lack of empirical constraints prior to \jwst\ meant that model predictions for the volume density of such sources spanned orders of magnitude, the observations lie at the very high end of these predictions \citep[e.g.][]{finkelstein23,Finkelstein.2024,harikane22,2023ApJ...951L...1P,Perez-Gonzalez.2025b}.
Since launch, \jwst\ spectroscopy has produced a high success rate for photometric selection of high-redshift sources and confirms the overabundance of high-redshift galaxies \citep[e.g.][and many others]{curtislake23, fujimoto23,arrabalharo23,Tang.2024,Tang.2025b,Naidu.2025}. Among the more likely physical mechanisms missing from the previous lower model predictions is a top-heavy initial mass function (IMF) -- proposed to become dominant at very low metallicities (where fragmentation into ``smaller'' stars would be less efficient, e.g., \citealt{Venditti2023}), as expected for galaxies at such early times \citep[e.g.,][]{Schneider2002,BL2003Nature,Cueto.2024}.  This possible mechanism is empirically testable via the detection of high ionization emission lines such as singly-ionized Helium \citep[\ion{He}{2} \lam1640, \ion{He}{2} \lam4687; e.g.,][]{bromm01,Schaerer2003,olivier22,maiolino23,nakajima22,Zackrisson.2024} and/or quadruply-ionized Neon ([\ion{Ne}{5}] \lam3426; e.g., \citealt{cleri23a,cleri23b,chisholm24}).

Additionally, early \jwst\ programs uncovered a surprising population of peculiar growing super-massive black holes (SMBHs, hereinafter active galactic nuclei; AGN) selected by broadened Balmer emission indicative of $>$1000 km s$^{-1}$ moving gas \citep[broad-line or BLAGN; e.g.,][]{ maiolino23, harikane23, kocevski23, larson23,matthee24}. From X-ray to Radio, this population appears to lack most multi-wavelength AGN signatures; their estimated number densities and black hole masses challenge expectations built on pre-\jwst\ predictions \citep{ananna24, yue24, lupi24, lambrides24b, gloudemans25, williams24, wang25, akins25, ronayne25}. This includes a subset of sources, dubbed “Little Red Dots” or LRDs, which have launched intensive debate surrounding their formation and fate in attempting to physically motivate their observed ``v-shaped'' spectral energy distributions (SEDs) complete with a restframe UV excess alongside red \jwst/NIRCam colors \citep[$m_{277} - m_{444} > 1.5$ mag;][]{barro24,furtak24, matthee24, greene23, kokorev23,  kocevski23, kocevski25, kokorev24, 2024ApJ...968....4P,akins25}. 
Some attempts to explain the discrepancy between canonical AGN characteristics and these new sources are centered on 1) variance in the dust properties \citep[i.e., temperatures, grain sizes;][]{casey24, setton25}, 2) the possibility for more extreme accretion \citep[e.g.,][]{king24, pacuccinarayan24, yue24, lambrides24b, madau24, lupi24}, or 3) or extremely compact star formation de-emphasizing the need for an AGN at all \citep[e.g.,][]{2024ApJ...968....4P,williams24, akins25, leung24, baggen24, wang25, whalen25}. The majority of spectroscopically-confirmed \jwst\ BLAGN samples (including LRDs) are from relatively shallow observations (${<}3$ hr, in \jwst\ cycles 1 \& 2), targeted with lower resolution \jwst/NIRSpec prism (R$\sim$100) spectroscopy. While these essential early surveys were critical for confirming photometrically-selected samples, breaking the degeneracy between competing interpretations of these sources requires deeper and higher-resolution data. Specifically, reaching depths that allow for robust searches of faint ionization lines ($F_{line} \simeq 0.1 \times F_{[OIII]}$)  will aid in understanding the complete ionization state of the hot ionized gas, and higher-resolution (R$\gtrsim$1000) will enable better disentangling of complex blended line systems such as what has been observed in the Balmer series in BLAGN with \jwst.

Another curious result from early \jwst\ papers was the discovery of galaxies with very large presumed stellar masses (log$(M/M_\odot){>}10$) at redshifts up to $z \sim 9$ \citep[e.g.,]{labbe22}.  If these galaxies were as massive and abundant as initially reported, they would directly challenge predictions from $\Lambda$CDM \citep[e.g., see early discussion in][]{mbk22}.  Updated results from multiple studies indicate the masses are likely somewhat lower \citep[e.g.,][]{kocevski23, greene23, chworowsky23, leung24,2024ApJ...968....4P}.  With a modest area coverage of $\sim$90 arcmin$^2$, the CEERS survey \citep{Finkelstein.2024} appears to contain $\sim$50 such systems at $z > 4$, with $\sim$20 at $z > 6$, some even appearing to have ceased their star-forming activity \citep[e.g.,][]{carnall23, looser23}.
While growing to log$(M/M_\odot){>}10$ in $<$1 Gyr is not impossible, it is not predicted by current models at the observed abundance \citep{Valentino.2023,Weibel.2025,Wang.2025}.  Such galaxies must form their stars at a remarkably rapid rate \citep[e.g.,][]{Glazebrook.2024,EuclidCollaboration.2025}. Those that are already quiescent at high redshifts must have shut down their star formation just as rapidly \citep[e.g.,][]{Carnall.2023,Carnall.2024}.  The best path to understanding the formation of these massive galaxies is to reveal their star-forming histories \citep[SFHs; e.g.,][Khullar et al., in prep; also see \citealt{Matthee.2025} for a review]{chworowsky23,Carnall.2024,Glazebrook.2024,Xiao.2024}.

Addressing these three questions --- 1) the presence of a top-heavy IMF in the early galaxies, 2) the nature of the gas surrounding high-z BLAGN and LRDs, and 3) the SFHs of massive early galaxies --- requires deep, rest-frame optical spectroscopy of targeted high-redshift sources. The High-[Redshift+Ionization] Line Search (THRILS; PIs T. Hutchison \& R. Larson, GO-5507) program obtained deep ($>$8 hr) \jwst/NIRSpec spectroscopy with the G395M medium-resolution (R$\sim$1000) grating for 184 sources in the EGS field (with the majority of these sources previously photometrically discovered in the Cosmic Evolution Early Release Science, or CEERS, Survey \jwst/NIRCam imaging,  \citealt{bagley22b,finkelstein25}). 
In this paper, we present the THRILS survey design (\S \ref{sec:surveydesign}), describe the observations and data reduction (\S \ref{sec:data}), and publish a spectroscopic redshift catalog including a list of non-detections (\S \ref{sec:catalog}). Finally, we detail an overview of the THRILS program science goals (\S \ref{sec:sciencegoals}). 
In this work, we assume a standard flat cold dark matter universe with a cosmological constant $(\Lambda$CDM), corresponding to the nine-year Wilkinson Microwave Anisotropy Probe (WMAP9, \citealt{Hinshaw.2013}).

\section{Survey Design}\label{sec:surveydesign}

The EGS / CEERS \citep{finkelstein25} field has been heavily studied by teams around the globe, with growing excitement regarding spectroscopic detections of galaxy overdensities from galaxies at various epochs in the early Universe \cite[z = 4.9, 7.8, 8.7; e.g.,][]{larson22, tang23, whitler22, tilvi20, jung22, haro23, whitler25}.  Such galaxy overdensities highlight the thrilling nature of the CEERS field and its scientific potential to detect even fainter sources in these same early epochs.
Additionally, the deep and extensive multi-wavelength photometry of the broader legacy EGS field enables high-fidelity photometric redshifts, and the plethora of public spectroscopic data verifies many of our photometric selections. 

\subsection{Ancillary Data Catalog Creation}{\label{sec:catalogcreation}}

The primary science goals of THRILS require photometric selection of galaxies across $4\lesssim z \lesssim 10$.  To do this, we created a combined catalog for the EGS field using the public CEERS NIRCam and MIRI imaging \citep{finkelstein25,bagley22b,yang23}, and the CEERS NIRSpec and NIRCam WFSS spectroscopy (\citealt{pirzkal17}; Arrabal Haro et al. in prep).  Additionally, for legacy ancillary data we include products from CANDELS \citep{grogin11,koekemoer11}, specifically the public CANDELS \hst+\spitzer\ catalog \citep{stefanon17} and the CANDELS EGS catalog from \citet[hereafter F22]{finkelstein22}. We also use external \jwst\ products from the Dawn \jwst\ Archive\footnote[1]{\href{https://dawn-cph.github.io/dja/spectroscopy/nirspec/}{https://dawn-cph.github.io/dja/spectroscopy/nirspec/}} (DJA; \citealt{degraaff24, heintz24}) for recent spectroscopic redshifts, as well as a compilation of ground-based spectroscopic programs \citep{mosdef, larson22, jung18, urbanostawinski24}. For the CEERS imaging, we used the UNICORN photometric catalog (S. Finkelstein et al., in prep.), which places particular emphasis on measuring accurate colors and total fluxes and includes photometric redshifts with improved templates \citep{larson22b}. 

\begin{figure*}[!ht]
    \centering
    \includegraphics[width=0.95\linewidth,trim={4mm 3.65cm 4mm 2.7cm},clip]{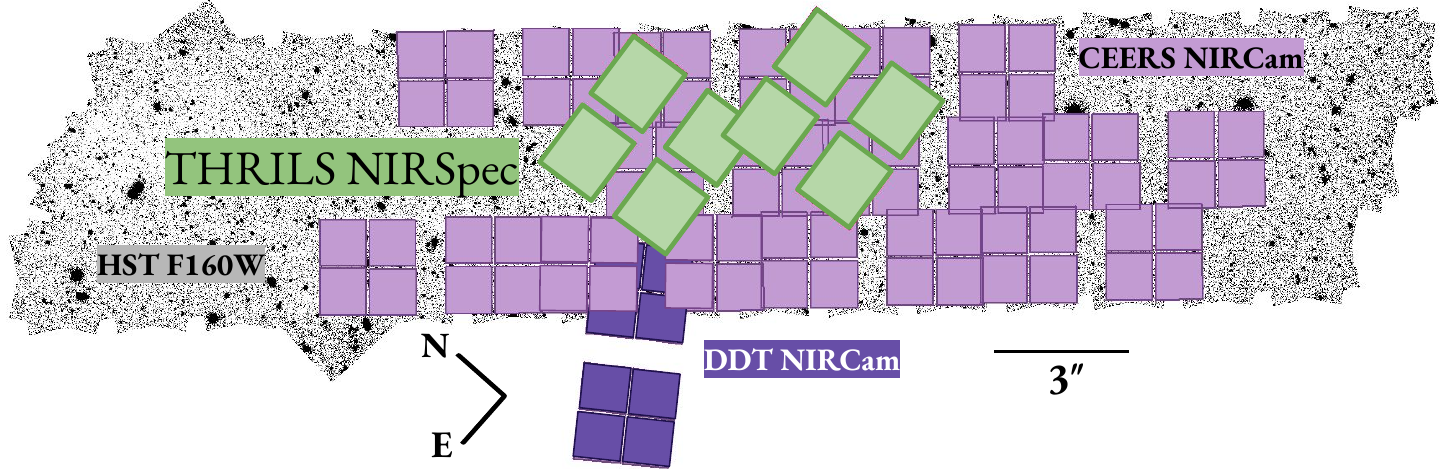}
    \caption{The two THRILS NIRSpec pointings (green) overlaid on the \hst/F160W mosaic image from CANDELS \citep{grogin11,koekemoer11}, with the CEERS NIRCam coverage outlined in purple \citep{bagley22b, finkelstein25}. The parallel NIRCam imaging from NIRSpec DDT program 2750 \citep{haro23} is shown in dark purple. }
    \label{fig:fov}
\end{figure*}

To create a comprehensive combined catalog, we match all source catalogs to within 0.5'' using TOPCAT \citep{topcat}. This yielded 169,011 unique sources across all of EGS. We then created several columns for our use: RA, DEC, a magnitude in three bands (H, 3.6 $\mu$m, and 4.5 $\mu$m), a photometric redshift ($z_{phot}$), and a spectroscopic redshift ($z_{spec}$). We also created columns labeled $z_{phot} $ vetted and $z_{spec}$ vetted for each source. We populated these columns with values in order of most recent or highest resolution. For the RA and DEC columns, we use the coordinates from NIRCam where available (including the WFSS data), then from NIRSpec, followed by the CANDELS \hst\ coordinates for those sources that were not observed with NIRCam or NIRSpec (corrected to match the more accurate, Gaia-anchored astrometric reference frame for the CEERS NIRCam image mosaics), and finally the MIRI coordinates for the $\sim$840 sources that were only detected in MIRI imaging.

To fill out the magnitude columns, we chose three representative bands: H, $3.6\mu$m, and $4.5\mu$m, as there are observations at these wavelengths from both \jwst\ and \hst\ or \spitzer\ -- the latter two of which overlap with the NIRSpec G395M filter. For the H-band magnitudes ($m_H$), we use the measured flux in the \jwst/NIRCam F150W filter where observed and the \hst/WFC3 F160W for those outside of the NIRCam coverage. For any sources that do not have an observed magnitude in either of these bands (i.e., those that are only MIRI detected), we use a value of $-1$ in this column. We follow a similar procedure for the other two magnitude bins ($m_{3.6}$ and $m_{4.5}$), using the \jwst/NIRCam F356W (including from the CEERS WFSS data) or F444W, where observed, or the \spitzer\ Ch1 or Ch2 when outside of the \jwst\ coverage. Again, for any sources that do not have an observation in one of these bands, we assign a $-1$ value to that column. 

The photometric redshift column ($z_{phot}$) is populated by first using the redshift values based upon the CEERS NIRCam observations \citep{cox25}, then using those from the \hst+\spitzer\ CANDELS catalog \citep{kodra23}, and finally those from the CANDELS EGS catalog (F22). Each of these catalogs has a ranking value for its photometric redshifts, which we use to assess the reliability of the measurements and assign a grade to the $z_{phot}$ vetted column. Sources from the NIRCam and the F22 catalog were vetted for photometric redshifts by meeting a set of criteria, namely requiring detection significance in at least two photometric bands and reasonably well-constrained photometric redshifts. The CANDELS catalog values were taken at face value unless they had bad flags, such as those near stars or at the edges of the images \citep{stefanon17}. Photometric redshifts are vetted by applying a value of 2 if the $z_{phot}$ is believed, a 1 if it is likely but uncertain, a 0 if it is NOT believed, and a $-1$ if it is not vetted. This means that sources with a value in the $z_{phot}$ column and a 2 in the $z_{phot}$ vetted column are likely robust measurements; however, those with a 0 in the vetted column should not be used in any analysis.

The spectroscopic redshift column ($z_{spec}$) first uses the value in the CEERS spectroscopic redshift catalog (Arrabal Haro et al. in prep), then the value from the DJA, followed by the value from the CANDELS catalog, followed by the ground-based redshifts from Keck/MOSFIRE (MOSDEF; \citealt{mosdef}) and DEIMOS \citep{urbanostawinski24}. Each of these catalogs assigns a grade to the believability of the redshift measurement, and we use those grades to determine a value for our vetting. Spectroscopic redshifts are vetted by applying a value of 2 if the $z_{spec}$ is believed, a 1 if it is likely but uncertain, a 0 if it is NOT believed, and a $-1$ if it is not vetted. For example, those sources in the DJA that have a grade of 3 (flagged by DJA as robust) are assigned a $z_{spec}$ vetted value of 2, but those that have a DJA grade of 0 (flagged by DJA as a data quality issue) or 1 (flagged by DJA as having no features) are assigned a vetted value of 0 and not used in the target selection.

\subsection{Target Selection}{\label{sec:targetselection}}

We selected targets for our observations based upon the three main science goals of the program: 1) higher-redshift ($z{>}7$) and high ionization lines, 2) high-redshift AGN ($z{>}4$), and 3) high-redshift massive galaxies ($z{>}4$), as described below. 

We selected sources for Science Case 1 (higher-redshift and high-ionization galaxies, see \S \ref{subsec:SC1}) using several methods. To identify sources that met both criteria, we selected sources with spectroscopic or photometric redshifts at $z>6$, which were expected to have higher ionization based on photometric color selection techniques (potentially powered by recent starbursts). We summarize the method here: in brief, we used several photoionization models of young ($<$50 Myr), sub-Solar-metallicity star-forming galaxies spanning a broad range of ionization.  We generated the NIRCam colors F356W--F444W as the initial color cut, selecting photometric redshifts for which strong lines, driven by very high ionization, dominate the color.  Secondary selections were made using NIRCam colors F277W--F356W and F410M--F470N, following similar criteria.  The final rankings were made from a combination of the primary and secondary color selections.

Separately, to build a larger sample of general high-redshift sources, we followed the selection of \citet{finkelstein22c}, who selected sources at $z>8$ in the CEERS NIRCam imaging \citep{bagley22b} and visually vetted them as robust high-redshift candidates. 
Finally, for Science Case 1, we selected extreme emission-line galaxies (EELGs) in the CEERS imaging using the method described by \citet{davis24}. 

Our parent AGN sample for Science Case 2 (see \S\ref{subsec:SC2}) was compiled from several previously published catalogs spanning X-ray to radio wavelengths \citep{nandra15, park10, ivison07}, as well as recent \jwst\ MIRI imaging observations \citep{backhaus25}. We also included the broad-line AGN (BLAGN) from \citet{taylor24}, which were identified in the CEERS and  Red Unknowns: Bright Infrared Extragalactic Survey (RUBIES, GO-4233, PIs: A. de Graaff \& G. Brammer; \citealt{degraaff25}). We also include LRDs from \citet{Kocevski22} and \citet{barro24b}. 

The massive galaxies for Science Case 3 (see \S\ref{subsec:SC3}) were selected by running Dense Basis \citep{iyer21} on the CEERS NIRCam imaging to identify sources at $z>4$ with stellar masses log(M$_*/$M$_\odot) > 10$. We required that 95\% of the posterior on stellar mass lie at log(M$_*/$M$_\odot) > 10$, ensuring that our targets have a high likelihood of being massive galaxies. We then performed visual vetting on each candidate galaxy to remove spurious detections and sources with contaminated photometry, such as sources affected by blending with nearby objects, diffraction spikes, or NIRCam imaging artifacts. This selection process is similar to that of \citet{chworowsky24}.

\subsection{Weighting for MPT}

Using our combined catalog, we then need to assign weights for the MSA Planning Tool (MPT) to optimize our observations. We begin by applying a base weighting scheme that strongly favors higher redshifts and quickly downgrades sources that become too faint to be detected, regardless of their potential interest. The simplified formula would be:

$$\log(W) = \min(5,   z / 2)$$  
~\vspace{-10mm}\\
$$\textnormal{if\,} \min(m) \leq m_{threshold}\textnormal{, and}$$

~\vspace{-6mm}\\
$$\log(W) = \min(5, z / 2 + 2.5*[m_{threshold} - \min(m)])$$ 
~\vspace{-10mm}\\
$$\textnormal{if\,} \min(m) > m_{threshold},$$

~\vspace{-2mm}\\
where $W$ is the MSA weight, $z$ is the best redshift available for the object ($z_{spec}$ when available, otherwise $z_{phot}$), $m_{threshold}$ is the estimated detection depth of the program (here we use 28 ABmag), and $\min(m)$ is the magnitude in the brightest band ($m_{3.6}$ or $m_{4.5}$).
We down-weight sources with bad-pixel flags in the F22 catalog and those with star flags in the CANDELS catalog \citep{stefanon17}. To mitigate small sources at the edges of images that were receiving the maximum base MPT weights, we down-weighted all sources that did not have a vetted $z_{phot} > 2$ or a vetted $z_{spec} > 1$.

Using the MSA Planning Tool in the APT, we determine field centers by developing a source prioritization scheme, where the highest-priority sources are the spectroscopically confirmed ($z >$ 8) sources, the brightest spectroscopically confirmed LRD, and the massive ($z >$ 4) galaxies. Our other primary targets are those described in our target selection section above and are assigned the next-highest tier of priority. Lower priorities are assigned to progressively lower-redshift and fainter sources.  We further split our two pointings into two observing sequences each, in which we swapped out observed filler sources halfway through the observations ($\sim$4 hr per pointing) for new filler sources. 

\section{Data}\label{sec:data}

\subsection{Observations}{\label{sec:observations}}

We used 3-shutter slitlets and a 3-shutter nodding pattern in two different configurations per pointing, utilizing the NRSIRS2 rapid readout pattern for NIRSpec \citep{Boker.2023}. Our two pointings had slightly different total exposure times as the higher-priority and higher-redshift sources were predominantly located in Pointing 1 (Figure \ref{fig:fov}, left THRILS pointing). We used 18 groups per integration and 4 integrations per exposure in Pointing 1, thus setting the time per individual exposure to 5310.356 s. In Pointing 2 (Figure \ref{fig:fov}, right THRILS pointing), we used 17 groups per integration and 4 integrations per exposure, resulting in a single exposure time of 5018.5781 s. We observed 2 configurations per pointing, with 3 exposures per configuration. In our second configuration, we swapped out filler targets where possible but kept the primary targets in shutters for the full observation time. This yields a total exposure time of 8.85 hr (8.36 hr) for our primary sources in Pointing 1 (Pointing 2). The center RA \& Dec coordinates are (214.9578413 deg, +52.9193425 deg) for Pointing 1 and (214.8699359 deg, +52.8724952 deg) for Pointing 2. The two pointings were at slightly different position angles (PAs) with Pointing 1 at 3.9473 deg and Pointing 2 at 4.0115 deg. While the FOVs of the two NIRSpec pointings overlap in a small corner of both detectors, we did not place any sources in that region for either pointing due to mask optimization. Pointing 1 had 98 total targets, with 26 primary sources, 57 filler targets, and 16 contaminating sources. Pointing 2 covered 86 targets in total, with 33 primary sources, 44 filler targets, and 12 contaminants. In total, we observed 184 sources, of which 59 were from our three primary science cases and received the full exposure time.  We note that some ($\sim$24) of the sources observed in THRILS had previous spectroscopic measurements, which we discuss in \S\ref{sec:catalog}.

\subsection{Data Reduction}\label{subsec:reduction}
The THRILS NIRSpec data were reduced using the \jwst\ Calibration Pipeline version 1.17.1 (\citealt{bushouse22}, DOI:10.5281/zenodo.7429939) with CRDS context jwst\_1350.pmap. The processing is based on standard pipeline parameters, with several modifications for specific steps. Here we give some relevant specifics, including deviations from standard processing: 

\textit{FFLAT Reference Files:} We use modified FFLAT reference files. The current reference file for the FFLAT (Fore optics FLAT-field) contains very large values in the error extension, resulting from the observed calibration star not matching the expected model during commissioning observations. Only some pixels (wavelengths) have these very large error values, which then propagate to produce unreasonably large flux errors in the reduced spectra. As a result, we mask out the entire error array from the FFLAT files for our reductions.

\textit{Stuck-Closed Shutters:} Unexpectedly stuck-closed shutters are detected through the identification of missing sky emission in the background-unsubtracted 2D spectra. The MSA metafiles are modified to account for these failed closed shutters. The dithering nods in which the target is behind a failed closed shutter are omitted when combining the dithers to generate the final spectra. This is reflected in the measured total exposure time per source in the catalog (\S \ref{sec:catalog}). The MSA metafiles are also modified to implement the correct extraction at the expected location for known companion sources.

\textit{Point Sources:} The pipeline was instructed to treat all targets as point sources (SRCTYPE = ``POINT'') -- such that no BARSHADOW correction was applied, and the 1D spectra are in units of flux density rather than surface brightness. 
The default pipeline pathloss correction was employed, and the flux calibration used the default reference files for the adopted CRDS context. 

\textit{Optimized Extraction:} To obtain 1D spectra from each of these 2D spectra, we then use an optimal extraction \citep{horne86}, determined by spatially collapsing the 2D spectrum and fitting the spatial profile of the source with a Gaussian. We applied median filtering to remove any bright artifacts before collapsing the 2D spectrum. This also ensures that any potential contaminating objects are masked from the extraction profile for each source. The resulting spatial profile is then used as the weight for optimal 2D-to-1D extraction for each source, ensuring that the shape and extent of each source are correctly accounted for in the extraction weights.


\begin{deluxetable}{c|cc|cccc|ccc|cc}
\tablecolumns{12}
\tablecaption{The Abbreviated THRILS Survey Redshift Catalog}
\tablehead{
    \colhead{\textbf{THRILS}} & \multicolumn{2}{c}{\textbf{Coordinates}} & \multicolumn{4}{c}{\textbf{THRILS}} & \multicolumn{3}{c}{\textbf{Magnitudes}} & \multicolumn{2}{c}{\textbf{Prior Redshifts}}  \\ 
    \colhead{ID$_{MPT}$} & \colhead{RA} & \colhead{DEC} & \colhead{z$_{spec}$} & \colhead{Grade} & \colhead{ExpTime [s]} & \colhead{Category} & \colhead{H} & \colhead{3.60} & \colhead{4.50} & \colhead{z$_{phot}$} & \colhead{z$_{spec}$}} 
    \startdata
	102659 & 214.943146 & 52.94244 & 11.41 & 2 & 31862 & Primary & 28.4$\pm$0.1 & 27.95$\pm$0.04 & 27.68$\pm$0.04 & 11.6$^{10.0}_{10.0}$ & 11.42$^a$ \\
	117286 & 214.977181 & 52.926547 & 9.0307 & 1 & 15931 & Filler & 28.6$\pm$0.1 & 28.78$\pm$0.09 & 29.0$\pm$0.2 & 9.0$^{9.9}_{1.4}$ & -1 \\
	92641 & 214.823994 & 52.884186 & 8.93 & 1 & 15056 & Filler & 28.78$\pm$0.06 & 28.86$\pm$0.04 & 28.9$\pm$0.05 & 8.0$^{8.6}_{1.8}$ & -1 \\
	97442 & 214.838696 & 52.882226 & 8.929 & 3 & 30111 & Primary & 29.2$\pm$0.1 & 28.96$\pm$0.06 & 28.28$\pm$0.03 & 9.0$^{10.0}_{8.7}$ & -1 \\
	37717 & 214.93864 & 52.911749 & 8.7578 & 3 & 31862 & Primary & 26.48$\pm$0.02 & 26.76$\pm$0.01 & 26.24$\pm$0.009 & 9.0$^{9.0}_{8.8}$ & 8.768$^b$ \\
	19512 & 214.967531 & 52.932951 & 8.7146 & 3 & 31862 & Primary & 26.1$\pm$0.02 & 26.46$\pm$0.01 & 26.1$\pm$0.008 & 9.0$^{8.8}_{8.6}$ & 8.717$^b$ \\
    \enddata
\tablecomments{Abbreviated version of the THRILS Redshift Catalog.  The full version of this table is in the Appendix, and a machine-readable version is provided with the publication of this work. 
    Description of Columns: 1) ID in the THRILS MPT, 2-3) Coordinates, 4-5) Measured $z_{spec}$ in THRILS and grade for this redshift as described in \S\ref{sec:redshifts}, 6-7) Exposure time and target category for the source as described in \S\ref{sec:targetselection}, 8-10) Magnitude of the source in 3 representative bands as described in \S\ref{sec:catalogcreation}, 11-12) Prior measurements of $z_{phot}$ and $z_{spec}$ from ancillary data as described in \S\ref{sec:catalogcreation}.  \\
    ~\vspace{-3mm}\\
    {\footnotesize $^a$ \jwst\ DDT 2750 \citep{arrabalharo23}.\\
    $^b$ CEERS (Arrabal Haro et al. in prep).}}\label{tab:shortcatalog}
\end{deluxetable}


\section{Redshift Catalog}\label{sec:catalog}

Here, we provide a spectroscopic redshift catalog for all sources with detections in the THRILS survey. Included columns are the MPT ID from THRILS, RA and Dec of the source, a measured spectroscopic redshift ($z_{spec}$), a redshift grade, exposure time on source, target category (Primary, Filler, or Contaminant), magnitudes in 3 bands (m$_H$, m$_{3.6}$, and m$_{4.5}$), and prior photometric ($z_{phot}$) and spectroscopic ($z_{spec}$) redshift information as described above in \S\ref{sec:catalogcreation}. Sources that have \hst-only derived $z_{phot}$ measurements, as they lie outside the CEERS NIRCam coverage, are marked with an asterisk (*). The first six rows are shown in Table \ref{tab:shortcatalog} as an example, and the whole catalog is provided in the Appendix (Table \ref{tab:redshiftcatalog}). A machine-readable table is also provided with the publication of this paper.

\begin{figure}
    \centering
    \includegraphics[width=\linewidth]{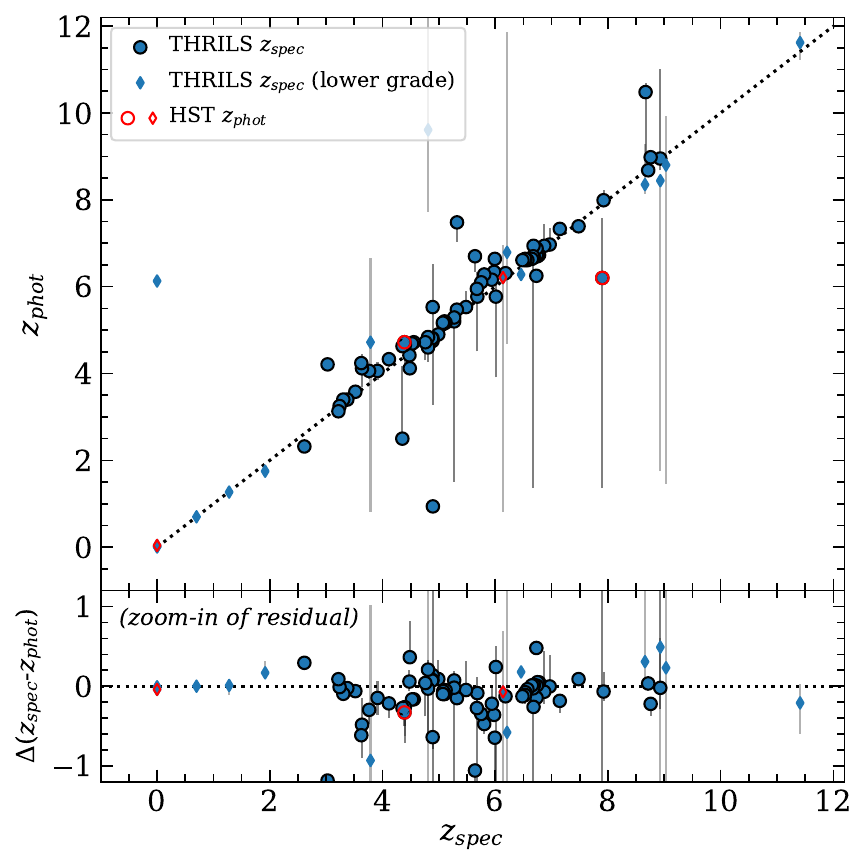}
    \caption{\textbf{Spectroscopic vs photometric redshift.} (\textit{top}) $z_{spec}$ vs $z_{phot}$ for all 89 sources with measured redshifts in THRILS. The robust redshift measurements (grade = 3) are shown as blue circles, while the lower confidence measurements (grade $<3$) are shown as diamonds.  We annotate all $z_{phot}$ derived from \textit{HST}-only data with red borders. (\textit{bottom}) Residual of $z_{spec} - z_{phot}$, zoomed into sources around zero for clarity.}
    \label{fig:zspeczphot}
\end{figure}

\begin{figure*}[]
\centering
\includegraphics[width=\linewidth]{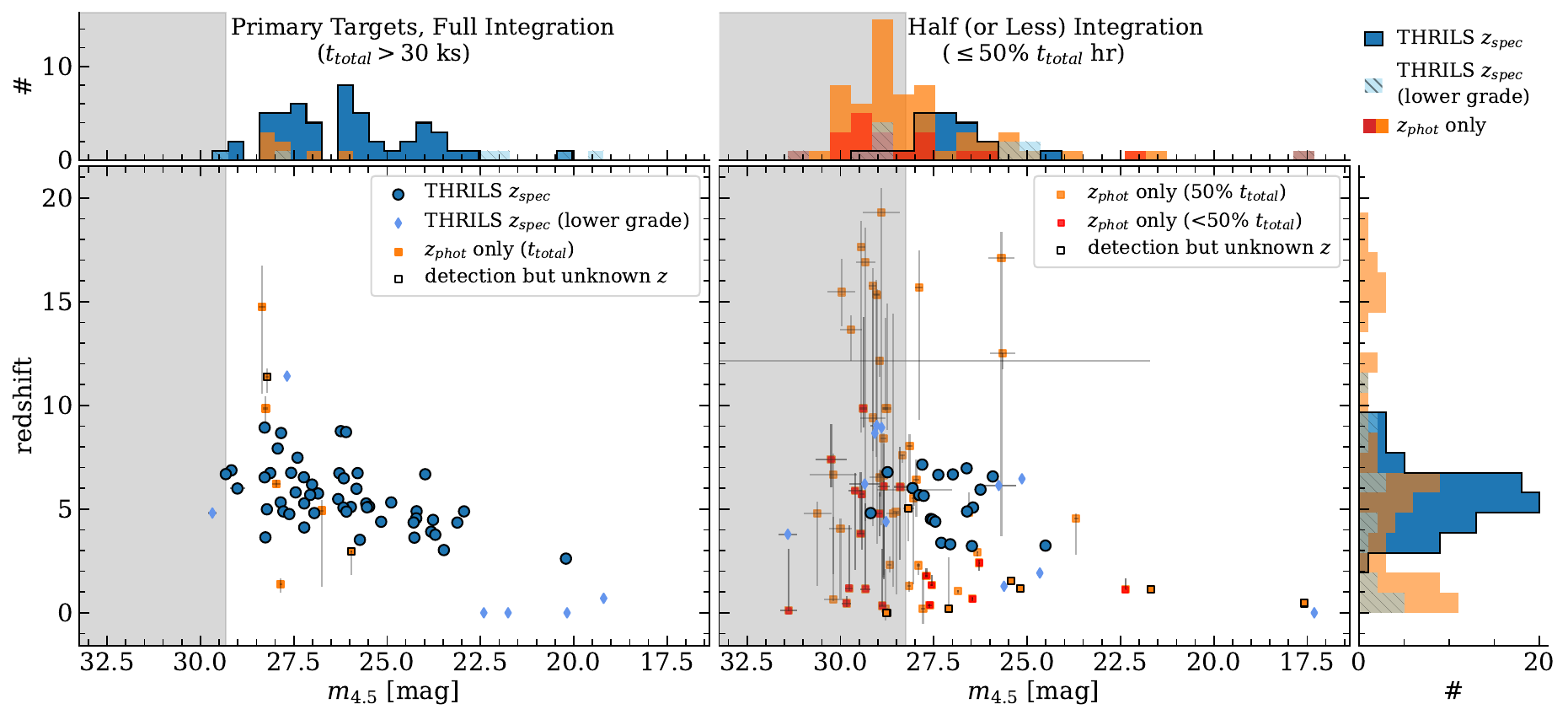}
\caption{
    \textbf{Detections and magnitude depths.}
    Plot of the source magnitude at 4.5 $\mu$m from \jwst/NIRCam F444W (or \spitzer\ Ch2 when not observed with \jwst)  vs redshift for the THRILS survey. 
    (\textit{left}) The THRILS primary science targets, with full integration time for each pointing ($>$30 ks). The THRILS spectroscopic redshifts ($z_{spec}$) are shown in blue, with high confidence (circles, grade = 3) and lower confidence (diamonds, grade = 1--2) redshifts.  For the primary targets that yielded non-detections (orange squares), we use their photometric redshift ($z_{phot}$) as the displayed redshift.
    (\textit{right}) The THRILS filler targets or contaminating sources, which had half the integration time (or less) per pointing.  The THRILS spectroscopic redshifts are again shown in blue. The non-detected sources are separated into half integration (orange, $\sim$15 ks; three dithers) and less than half (red, $<$12 ks; one or two dithers only).
    The grey shaded regions show the NIRSpec detection threshold for non-EELGs in both panels, highlighting the depth achieved at both exposure times and advocating for more deep NIRSpec programs. 
    }
    \label{fig:threshold}
\end{figure*}


\begin{deluxetable*}{c|cc|cc|ccc|cc}
\centerwidetable
\tablecolumns{10}
\tablecaption{Abbreviated Non-Detections from the THRILS Survey}\label{tab:shortnondetections}
\tablehead{
    \colhead{\textbf{THRILS}} & \multicolumn{2}{c}{\textbf{Coordinates}} & \multicolumn{2}{c}{\textbf{THRILS}} & \multicolumn{3}{c}{\textbf{Magnitudes}} & \multicolumn{2}{c}{\textbf{Prior Redshifts}}  \\ 
    \colhead{ID$_{MPT}$} & \colhead{RA} & \colhead{DEC} & \colhead{ExpTime [s]} & \colhead{Category} & \colhead{H} & \colhead{3.60} & \colhead{4.50} & \colhead{z$_{phot}$} & \colhead{z$_{spec}$}} 
    \startdata
	127239 & 214.85961 & 52.858029 & 15056 & Filler & -1 & -1 & -1 & -1 & 6.0611$^a$ \\
	136697 & 215.003204 & 52.896604 & 15931 & Filler & 23.5$\pm$0.003 & -1 & -1 & 5.0$^{5.8}_{5.4}$* & -1 \\
	139799 & 214.948036 & 52.926073 & 10621 & Contaminant & 25.7$\pm$0.07 & -1 & -1 & 2.0$^{8.0}_{1.4}$* & -1 \\
	160204 & 214.841267 & 52.891728 & 15056 & Filler & 27.17$\pm$0.07 & -1 & -1 & 5.0$^{5.4}_{2.5}$* & -1 \\
	139366 & 214.9154 & 52.904181 & 15931 & Filler & 27.5$\pm$0.1 & -1 & -1 & 10.5$^{10.0}_{3.2}$* & -1 \\
	160727 & 214.913099 & 52.941113 & 15931 & Filler & 27.7$\pm$0.1 & -1 & -1 & 3.0$^{10.0}_{2.2}$* & -1 \\
    \enddata
\tablecomments{Abbreviated version of the table of non-detections in the THRILS data sorted in order of brightness.  We list the full non-detection table in the Appendix Table \ref{tab:shortnondetections}. 
    Description of Columns: 1) ID in the THRILS MPT, 2-3) Coordinates, 4-5) Exposure time and target category for the source as described in \S\ref{sec:targetselection}, 6-8) Magnitude of the source in 3 representative bands as described in \S\ref{sec:catalogcreation}, 9-10) Prior measurements of $z_{phot}$ and $z_{spec}$ from ancillary data as described in \S\ref{sec:catalogcreation}. \\
    ~\vspace{-2.5mm}\\
    {\footnotesize * Sources with $z_{phot}$ values from \hst\ only, they do not have \jwst/NIRCam coverage. \\
    $^a$ DJA \citep{degraaff24, heintz24}.} }
\end{deluxetable*} 


\subsection{Redshift Measurements}\label{sec:redshifts}

The redshifts were measured using an automated line-finding software \citep{larson18,larson23} designed to fit and identify emission-line features. The code uses a photometric redshift prior to determine the expected brightest line in the spectrum (i.e., \oiii\ \lam5008 or \ha\ \lam6564), and fits the redshift to the line center. At least two team members visually vetted all redshifts published in this work for accuracy. For those with a single detected emission line, the photo-z was used to obtain best-guess redshifts. When line identification is ambiguous, the redshift grade is lowered (see below). For the few sources with detected emission that could not be identified, we assigned a $z_{spec} = -2$. For sources with detected continuum but no discernible emission with which to measure a redshift, we assigned a $z_{spec} = -3$. 

The spectroscopic redshifts are graded as follows: 3 denotes a robust spectroscopic redshift, with multiple or distinct emission lines detected. A grade of 2 is assigned when faint emission lines are detected but are likely to be correctly identified, and thus the redshift is considered secure. A grade of 1 is for those with tentative line detections, but which are in agreement with prior photometric or spectroscopic redshift measurements, or those with one line detected, and an approximation is made for the redshift. A grade of 0 is assigned to sources with detected lines that cannot be identified ($z_{spec} = -2$) or with detected continuum but no emission lines detected ($z_{spec} = -3$). Finally, there were a small number of sources that we suspect are stars or brown dwarfs and were thus assigned $z_{spec} = 0$ with a grade of 1. 

In total, we publish spectroscopic redshift measurements for 89 sources in THRILS and an additional 13 sources that show emission lines or continuum but for which we could not discern a likely redshift.  
In Figure \ref{fig:zspeczphot} we show the measured $z_{spec}$ vs $z_{phot}$ for these sources.  For clarity, we distinguish between those with robust measurements (grade = 3, circles) and those with lower-confidence redshifts (grade $<3$, diamonds). We also highlight sources with no \jwst/NIRCam imaging (and thus only with a $z_{phot}$ from \hst) with red outlines. In the bottom panel of Figure \ref{fig:zspeczphot}, we show a zoom in on the residuals around zero, highlighting that most of our photometric redshifts are in good agreement with the measured spectroscopic redshifts. 

We present an abbreviated view of the catalog of these sources in Table \ref{tab:shortcatalog} and provide the full redshift catalog in Table \ref{tab:redshiftcatalog} in the Appendix. Of the 89 sources with spectroscopic redshifts derived from the THRILS program, only 24 had previous spectroscopic measurements (primarily from NIRSpec/MSA prism observations).  We note any previous spectroscopic redshift measurements in the final column of both Tables \ref{tab:shortcatalog} \& \ref{tab:redshiftcatalog}, with table footnotes denoting the reference catalogs or publications from which the previous measurements were obtained.

\subsection{Discussion of Non-Detections}\label{sec:nondetections}

In the THRILS data, 81 sources were not detected (i.e., those with no detectable signal), of which only 2 were primary science targets, and only 6 had the complete $>$8 hr exposure time depth. A majority of the non-detected sources were faint filler targets and/or contaminating sources that received $\leq$50\% of the program's full exposure time.
The abbreviated view of the non-detected source list is shown in Table \ref{tab:shortnondetections}, and we include the full non-detection catalog in Table \ref{tab:nondetections} in the Appendix. The format follows the same as that of the redshift catalog in \S \ref{sec:catalog}, however without a redshift measurement or grade. The non-detected sources are sorted in order of brightness in descending wavelength order, with the brightest first. Sources that either lack a detection or are not observed in a specific photometric band are marked with a $-1$ in that column, and we include these sources at the top of the catalog.

The detection limits of our program are shown in Figure \ref{fig:threshold}, with spectroscopically detected and non-detected ($z_{phot}$ only) THRILS sources vs 4.5\micron\ magnitude ($m_{4.5}$).  The left panel includes all sources at full depth for each pointing ($>$30ks or > 8 hr), with the robust THRILS redshifts (grade = 3) shown as blue circles and the lower-confidence THRILS redshifts (grade = 1--2) as light blue diamonds. The right panel shows sources at half depth or less, with sources that were changed out after 4 hr in each pointing (orange, $\sim$15ks) or sources that were contaminants, appearing in only one or two dithers (red, $\leq$12ks).  Sources with no spectroscopic redshift assigned are shown by their $z_{phot}$ (orange and red squares).
Above each panel, we show the distribution of $m_{4.5}$ for the robust THRILS spectroscopic redshifts (blue with black border), the lower confidence THRILS redshifts (light blue with black hatching), and the photometric-only redshifts (i.e., non-detections, orange and red).

In each panel, we highlight the non-detection threshold in gray-shaded regions at both the full survey depth (left) and half the program depth (right). The shaded area is drawn at the magnitude of the faintest robustly-detected source, $m_{4.5} = 29.3$ mag, for the full survey depth. For the half-depth, we find a detection threshold of $m_{4.5} \simeq 28.25$ mag, nearly 1 dex shallower than the $>8$ hr full-integration time. We note that the two secure THRILS detections within the shaded region in the half-depth panel are unique, in that they are two EELGs whose very strong nebular lines are the reason they could be detected at such faint NIR magnitudes.\\

~\vspace{-1.2cm}\\
\section{Survey Science Goals}\label{sec:sciencegoals}
In the following sections, we outline the three key science cases that drove this observing program.  As part of a broader discussion about the need for still-deeper NIR spectroscopy, we include example spectroscopy from each science case to showcase the improvements in sensitivity of the THRILS program ($>$8 hr) compared to other programs in the EGS field that have had $<$1 hr of integration  in this same filter and resolution (G395M),\footnote[2]{We note that these programs were intended to cover wide areas of their target fields, therefore their observing strategies were to use shallower integrations to accommodate a larger coverage on the sky. The comparison of these programs to THRILS is not to devalue these previous programs, but rather to emphasize the importance of deeper follow-up observations.} such as CEERS (Arrabal Haro et al. in prep) and RUBIES \citep{degraaff25}.

\subsection{Science Case 1: Probing the Nature \\ of the IMF at Cosmic Dawn}\label{subsec:SC1}

Among the many predictions for the earliest galaxies is the notion that they should contain extremely low-metallicity (perhaps even primordial) stellar populations \citep[e.g.,][]{jaacks18,LiuBromm2020,Venditti2023}.  Such systems are expected to have a very different stellar IMF than the present-day Universe \citep[e.g.,][]{Bromm2013,Klessen2023}, with a ``top-heavy" IMF (higher abundance of more massive stars) predicted to become dominant at these early times due to a combination of low metallicities and a rising CMB temperature floor \citep[e.g.,][]{larson98,bromm01Zcrit,clarke03,SchneiderOmukai2010,chon21,sharda22}.   
While it is unclear whether we should expect to see truly metal-free stars with \jwst, these observations now probe $<$300 Myr after the Big Bang \citep[e.g.,][]{Naidu.2025}. It would be surprising if \jwst\ observations did not reveal that the IMF began to evolve early.  Top-heavy stellar populations have a ratio of UV luminosity to unit stellar mass that rises with the dominant stellar temperature.   This lack of evolution is precisely what one would predict if a top-heavy IMF dominated the light from such early galaxies.  In fact, \citet{finkelstein23} found that the observed galaxy population was consistent with the expected evolution if the $z >$ 10 galaxies had their UV luminosities boosted by factors of $\sim$2$\times$ (see also \citealt{Yung_IMF2024}), consistent with expectations from extremely metal-poor systems \citep[XMP; $Z < 3$\% $Z_{\odot}$; e.g.,][]{raiter10,berg21}.   
The advanced spectroscopic capabilities of \jwst\ make these scenarios immediately testable, as stellar populations dominated by XMP (or zero-metallicity) stars would exhibit extremely high-ionization emission lines \citep[e.g.,][]{pawlik11,nakajima22}.

\begin{figure*}[!ht]
\centering
\includegraphics[width=\linewidth]{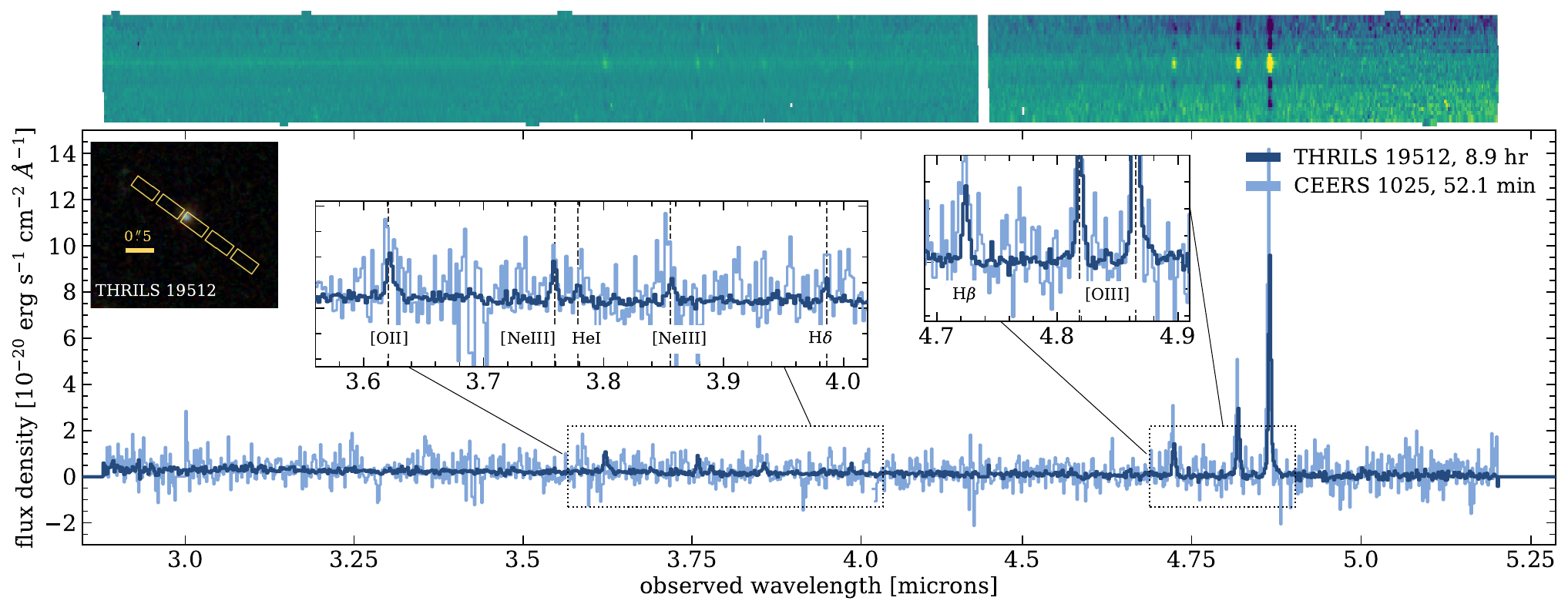}
\caption{\textbf{Spectrum of a source from THRILS Science Case 1.} THRILS-19512 is a $z=8.7146$ galaxy initially observed with NIRSpec by the CEERS program \citep[CEERS MPTID 1025;][]{Finkelstein.2024}. The 2D spectrum from THRILS is shown across the top, and the 7-filter color image from CEERS is shown in the inset to the left with the THRILS NIRSpec shutter overlaid. The publicly available 1D CEERS spectrum (light blue, Arrabal Haro et al. in prep) has an exposure time of 52.1 min, while the THRILS optimally extracted 1D spectrum (dark blue) has an exposure time of 8.85 hours. Inset plots highlight the detections of \hei\ \lam3890 and \hd\ \lam4103, which were not possible in the shallower data. }
\label{fig:sc1spec}
\end{figure*}


For the first science case, THRILS obtained deep NIRSpec/MSA spectroscopy of a sample of \jwst-discovered galaxies to probe their stellar populations.  The observing constraints and targets (described in \S\ref{sec:targetselection}) are optimized for both fainter sources and those that may host higher-ionization emission features.  The latter intended to determine if signatures of ultra-low metallicities or intermediate-mass (growing) black holes were present (for which low-redshift analogs from \citealt{cleri23b,olivier22,Berg.2019b} 
are valuable comparisons).  Very high ionization emission requires a strong ionizing continuum -- even in stellar populations dominated by extremely low metallicities coupled with a more ``typical'' IMF \citep[e.g.][]{kroupa13,chabrier03}, these spectral features would likely be weakly detected at best \cite[see modeling comparisons in][]{cleri25}.  Only in more extreme stellar population scenarios, i.e., XMPs prescribed by a top-heavy IMF \citep[such as those from Yggdrasil,][]{zackrisson2011}, would we expect a significantly strong detection of such high-ionization emission features.

A detailed study directly addressing the primary science in Science Goal 1 will be summarized in Hutchison et al. (in prep).
As a visualization of other science enabled for these high-z galaxies, here we showcase the improvements in sensitivity of the THRILS program ($>$8 hr) compared to other programs in the EGS field that have had $<$1 hr of integration time, such as CEERS (Arrabal Haro et al. in prep) and RUBIES \citep{degraaff25}. Figure \ref{fig:sc1spec} shows spectra for THRILS-19512, a high-redshift galaxy at $z=8.7146$ which CEERS first observed in the same grating/filter pair (G395M/F290LP; MPTID 1025). As a shallower spectroscopic program, CEERS had only $\sim$50 minutes of integration time (light blue, Arrabal Haro et al. in prep), compared to the THRILS (dark blue) 8.85 hr on this same source. In the deeper THRILS spectrum, fainter lines such as \hei\  \lam3890 and \hd\ are clearly detected, whereas in the shallower CEERS data, they fell below the noise threshold.  We note that in this figure, we use publicly available CEERS data, which was reduced using different processes and pipeline versions, as well as variations in slitloss corrections, likely accounting for any apparent differences in the detected line flux measurements. 
The increase in depth and sensitivity enabled by THRILS yields a significant boost in science from these data.  Additionally, and by design, more than half of the sources targeted as part of Science Case 1 in THRILS are fainter than $m_H$ (F150W) = 27.5 mag.
Detailed studies of faint early galaxies require the detection of the full suite of emission-line features, which is only possible with the sensitivity provided by observational depths similar to THRILS. See Hutchison et al. (in prep) and Davis et al. (in prep) for more details on the results from Science Case 1.

\begin{figure*}[!ht]
\centering
\includegraphics[width=\linewidth]{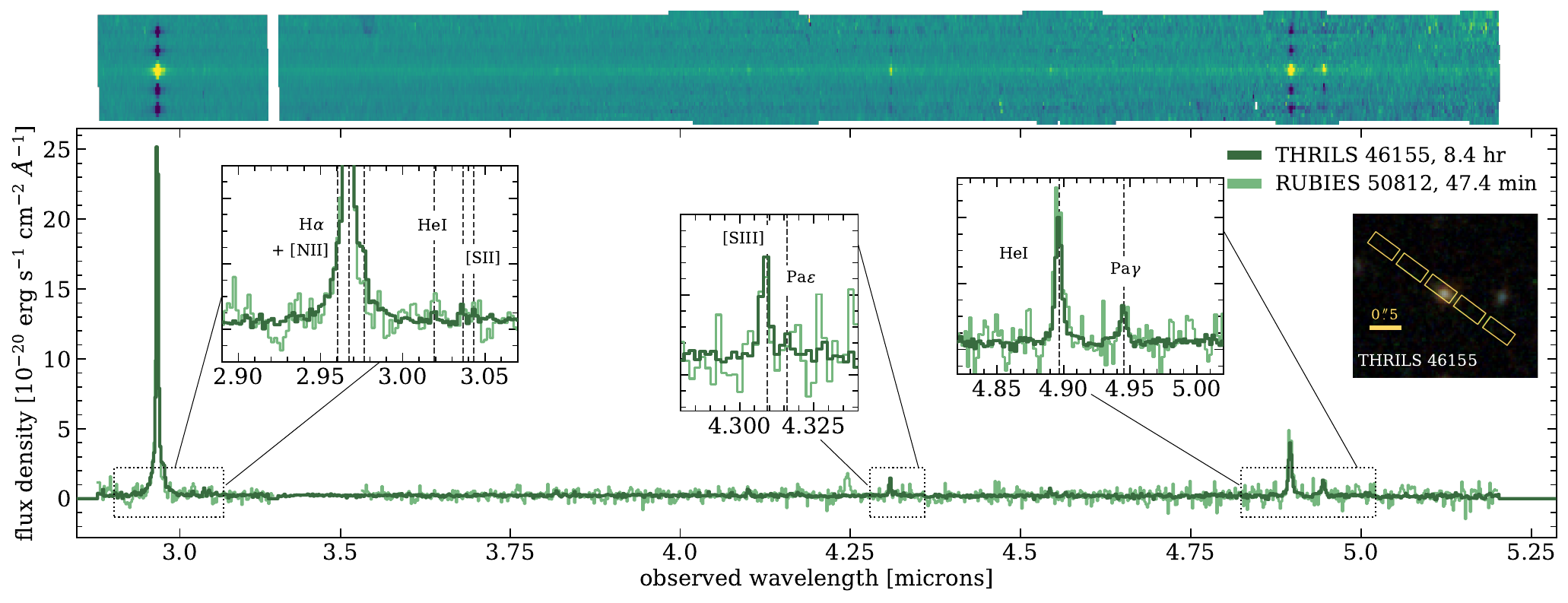}
\caption{\textbf{Spectrum of an AGN from THRILS Science Case 2.} THRILS-46155 is a BLAGN at $z=3.5179$ that was first observed for 47.7 min by the RUBIES program \citep[ID 50812;][]{degraaff25}. The 2D spectrum from THRILS (8.4 hr) is shown across the top, and the 7-filter color image from CEERS is shown in the inset to the right with the THRILS NIRSpec shutter overlaid. The publicly-available RUBIES data (light green) detects the broad component of \ha\ \lam6564 \citep{taylor24}, but the deeper THRILS data (dark green) is needed to measure the fainter \sii\ \lam\lam6718,6733 and \siii\ \lam\lam9069,9531 doublet lines shown in the insets.}
\label{fig:sc2spec}
\end{figure*}

\subsection{Science Case 2: Determining the Nature of \\ Puzzling Broad-Line AGN Discovered with \jwst}\label{subsec:SC2}

From ultra-dense gas to super-Eddington accretion, the origin, nature, and fate of newly-discovered Type 1 AGN candidates with \jwst\ remain one of the most contested areas of inquiry in early black hole-galaxy coevolution \citep{Inayoshi2025,lambrides24b,juodzbalis25, naidu25bhstar,lupi24}. The critical yet incomplete task of contextualizing high-z BLAGN, including LRDs, within the broader landscape of high-redshift galaxies has made it difficult to determine the intersection between extreme ISM properties and extreme accretion properties. For LRDs specifically, the common characteristics that all interpretations aim to account for are a)  peculiar ``v-shaped'' SEDs characterized by a puzzling UV excess alongside red \jwst/NIRCam colors ($m_{277} - m_{444} > 1.5$ mag, and an apparent Balmer break); b) broadened hydrogen Balmer emission ($>$1000 km s$^{-1}$); c) a general lack of multi-wavelength counterparts in the X-ray, sub-mm, and radio wavelengths; and d) a flat SED in the rest-frame NIR as inferred by \jwst/MIRI \citep[e.g.,][]{akins25,wang25, perezgonzalez24}. A subset of LRDs are found to have deep Balmer absorption within the Balmer emission features \citep[e.g.,][]{taylor25} and, for the handful of these sources observed to sufficient depths, rich iron emission \citep[e.g.][]{hviding25,lambrides25}. Part of the difficulty in concretely connecting the physical mechanisms that drive the rest-UV/optical spectral properties of both the new Type 1 BLAGN and LRD populations is the lack of spectral resolution and depth to sufficiently detail line emission on scales both close and far from the central emitting region. 

With THRILS, we are characterizing the density, temperature, and line widths of several BLAGN candidates (see Section \ref{sec:targetselection}). In addition to robustly measuring the kinematics of the broad-line region, we are also hunting for faint but important ISM-constraining lines (i.e., \nii\ \lam\lam6550,6585, \sii\ \lam\lam6718,6733, \oiii\ \lam4364) to compare and contrast the ionization state of the BLAGN populations discovered with \jwst. Here we highlight one of our AGN sources in Figure \ref{fig:sc2spec}, THRILS-46155, which is a BLAGN source first observed by RUBIES with the ID
50812 \citep{taylor24,degraaff25}. The THRILS $>$8 hr depth (dark green) enables the measurement of the \sii\ and \siii\ emission lines in this source, providing valuable metallicity-insensitive constraints on the ionization levels in this AGN \citep[e.g.,][]{Kewley.2019, Sanders.2020}. We note that in this figure, we use the publicly available RUBIES data, which was reduced using different processes and pipeline versions, as well as slitloss corrections, likely accounting for any apparent differences in the detected line flux measurements.

THRILS has already yielded significant surprises on this front, with the discovery of a rich series of faint iron features detected in a $z\sim7$ LRD \citep{lambrides25} 
previously spectroscopically observed by shallower programs \citep{Tang.2025,BWang.2025}. 
These lines uniquely constrain both the AGN nature of this source and provide new clues on the chemical enrichment history of LRD host galaxies. More details on the results from this Science Case will be presented in upcoming work by Davis et al. (in prep.) and Ganapathy et al. (in prep.).

\begin{figure*}[!ht]
\centering
\includegraphics[width=\linewidth]{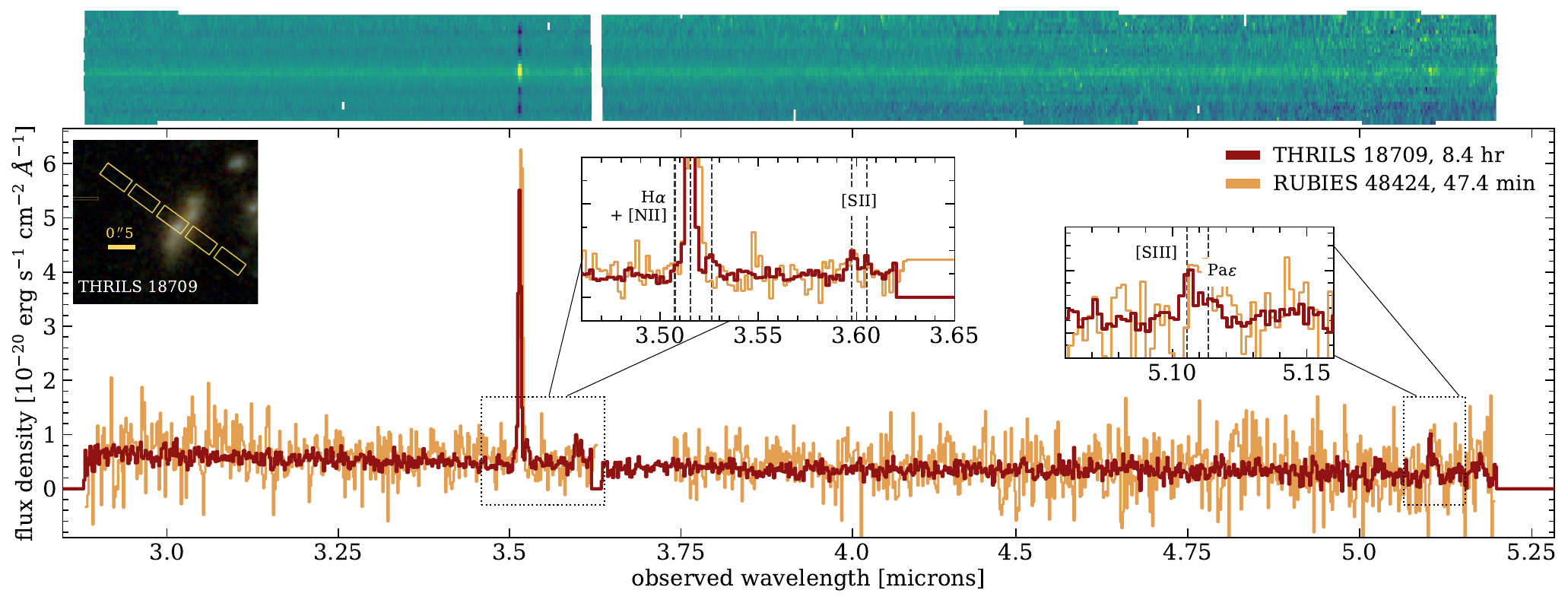}
\caption{\textbf{Spectrum of a Massive Star-Forming Galaxy from THRILS Science Case 3.} THRILS-18709 is a massive galaxy at $z=4.355$ which was first observed for 47.7 min by the RUBIES program \citep[ID 48424;][]{degraaff25}. The 2D spectrum from THRILS (8.4 hr) is shown across the top, and the 7-filter color image from CEERS is shown in the inset to the left with the THRILS NIRSpec shutter overlaid. The deeper THRILS spectrum (dark red) enables the disentangling of the \nii\ \lam\lam6550,6585 from the \ha\ \lam6564 line as shown in the inset. The THRILS spectrum also successfully measures the \sii\ \lam\lam6718,6733 doublet, which is within the noise of the publicly available RUBIES (light red) spectrum.}
\label{fig:sc3spec}
\end{figure*}

\subsection{Science Case 3: SFHs of massive galaxies at $z>4$}\label{subsec:SC3}

The emergence of massive galaxies in the early Universe has revealed that galaxies assembled stellar mass at earlier epochs than expected, with candidates selected from photometry with stellar mass log$(M/M_\odot)>10$ forming as early as $z\sim 9$ \citep[e.g.,][]{labbe22}. This population of candidate high-redshift massive galaxies is in non-negligible tension with current galaxy formation theories, as simulations have difficulty rapidly building up stellar masses to log$(M/M_\odot)>$10 in $<$1 Gyr \citep[e.g.,][]{steinhardt16, weaver22, lovell23, ferrara2023}. Photometry alone is insufficient to confirm the redshifts of these sources and recover their mass assembly histories with robustness. Age–dust–metallicity degeneracies make photometric SFH inferences largely prior-dominated (especially without a spectroscopic redshift; see discussion in e.g., \citealt{Perez-Gonzalez.2025}), and often different star formation history (SFH) models with varying degrees of flexibility can fit broadband photometric points equally well (\citealt{leja19, carnall19, chworowsky23}, and references therein).

NIRSpec enables spectroscopic confirmation and SFH constraints via emission-line and Balmer-absorption measurements, which are critical for detecting quiescence in galaxies and characterizing the oldest stellar populations in high-redshift galaxies.
%
The limited cosmic time available for the formation of these high-redshift galaxies means that their SFHs can be reliably recovered using complex, flexible, yet physically motivated non-parametric models. The flexibility of non-parametric models allows for the capture of rapid burst and quenching episodes seen at high redshifts (\citealt{looser23, endsley25, covelo2025}; studies based upon NIRCam photometry and/or NIRSpec prism spectra), while simultaneously avoiding biases of an assumed fixed SFH shape. Related to the latter point, there are established best practices in literature for avoiding biases and systematic uncertainties in inferred parameters from stellar population synthesis modeling, as well as mitigating the age-metallicity-dust degeneracy in spectroscopic modeling.  Such best practices include ensuring that multiple emission lines across a long wavelength arm are used for dust corrections, rest-frame UV to optical signatures are utilized to measure the age of stellar populations, etc. Many key spectral features can place limits on the star-formation/mass assembly and quenching timescales (e.g., instantaneous dust-unobscured star-formation rates and the ages of a diversity of stellar populations). Recent bursts of star formation are well traced by \ha\ (and to some extent \hb+\oiii), placing limits on star formation within the most recent 5-20 Myr. For galaxy populations that have ceased forming stars, the \hd\ absorption line (and other strong Balmer absorption lines) can indicate a post-starburst phase ($\sim$100 - 500 Myr after the initial burst) and the 3650\AA Balmer break (or D$_\textrm{n}$(4000)) places limits on older stellar populations of $\sim$300--500 Myr before observation \citep[e.g.,][]{Balogh99, Poggianti99,Poggianti09, Vergani10, Mansheim17}. 

THRILS is designed to constrain the mass assembly histories of these extreme objects and to further our understanding of their formation pathways by distinguishing among strong emission-line features, the Balmer break, and absorption features. Critically, the deblending of emission lines in proximity to each other -- \ha\ + \nii\ and \hb\ + \oiii\ -- is achieved in THRILS, and offers precise measurements of SFRs and physical conditions of the nebular regions within these systems (e.g., see \citealt{shapley2025} for state-of-the-art nebular line diagnostics in high-z galaxies calibrated to \jwst\ observations). While studies like \cite{covelo2025} acknowledge recent star formation in starburst/post-starburst galaxies with blended \ha\ + \nii\ measurements, THRILS is designed to definitively constrain SF at timescales of 10--20 Myr, and disentangle it from SF timescales of 100--200 Myr (via extant rest-frame UV measurements).

Here we show one of our massive galaxies from Science Case 3, THRILS-18709, first observed by RUBIES with ID 48424 \citep{degraaff25}. The THRILS spectrum (dark red) achieves the depth required to detect and disentangle the \nii\ emission from the \ha\ line profile, enabling a measurement of the metallicity in this source. Figure \ref{fig:sc3spec} also shows the successful detection of the \sii\ doublet and separation of \siii\ \lam9531 from \paE\ -- providing limits on the ionization levels in this source and further limiting the effect of degeneracies in the mass-weighted ages of stars and in gas phase metallicity. 

By combining spectroscopy with flexible, non-parametric SFH inference, THRILS will measure key formation timescales and burst durations, thereby confirming the presence of a diversity of stellar populations in massive galaxies in the early Universe, and simultaneously constraining how rapidly they assembled. These results will also directly inform whether star formation at high redshifts proceeds in a different regime than in lower-redshift galaxies. See Chworowsky et al. (in prep) for more results from Science Case 3.

\section{Conclusion}

In this paper, we summarize the THRILS survey goals, program execution, data reduction, redshift measurements, and detection threshold. We provide a redshift catalog for 89 THRILS sources with robust redshift measurements, visually vetted by the team. We also present the additional 81 sources that show no detectable signal in our data, despite the depth of our observations. The THRILS observational constraints yield NIRSpec detection thresholds of $m_{4.5\mu m}=29.3$ for an 8 hr integration and $m_{4.5\mu m}\simeq28.25$ for $\sim$4 hr, which can directly inform future deep spectroscopy surveys. We also highlight faint emission-line detections that are only possible with the deeper spectroscopy, underscoring the need for greater depth in future programs to answer key science questions across a wide range of science goals in the high-z Universe.

\begin{acknowledgments}
We acknowledge that a significant part of our work is done on stolen land, and we support the efforts of Land Back and true stewardship to those same peoples whose land is occupied. The PIs also note that we do not use the full name of \jwst\ due to the person after whom this telescope is named and their role as NASA administrator during the ``Lavender Scare'', as per the $\#RenameJWST$ protest movement.

The THRILS team would like to thank our program coordinators (PCs), Carol Rodriguez and Blair Porterfield, for their help with planning and executing our program. We would also like to thank Dan Coe, our Contact Scientist (CS), for his valuable insight and assistance with designing our MPT. 

We thank the programs that came before us, which enabled the cataloging of hundreds of thousands of galaxies in this field, using both ground- and space-based telescopes, both imaging and spectroscopy.  Such programs include CANDELS, MOSDEF, UVCANDELS, CEERS, RUBIES, and MEGA. 
We also thank efforts to create resources for the community, such as the Dawn \jwst\ Archive (DJA), which was used in this work to check previous \jwst\ redshifts for our catalog.

TAH's research is supported by an appointment to the NASA Postdoctoral Program at the NASA Goddard Space, administered by Oak Ridge Associated Universities under contract with NASA.
RLL appreciates support from a Giacconi Fellowship at the Space Telescope Science Institute, which is operated by the Association of Universities for Research in Astronomy, Inc., under NASA contracts NAS 5-26555 and NAS5-03127.
\end{acknowledgments}





\facilities{\jwst\ (NIRCam, NIRSpec, \& MIRI), HST (ACS, WFC3), Spitzer (Ch1, Ch2), Keck (MOSFIRE, DEIMOS)}

\software{astropy \citep{2013A&A...558A..33A,2018AJ....156..123A,2022ApJ...935..167A},  
          Cloudy \citep{2013RMxAA..49..137F}, 
          Source Extractor \citep{1996A&AS..117..393B}
          TOPCAT \citep{topcat},
          Dense Basis \citep{iyer21},
          IDL \citep{landsman93},
          \jwst\ Pipeline \citep{bushouse22},
          SAOImage DS9 \citep{SmithsonianAstrophysicalObservatory.2000}}


\appendix

Here we provide a spectroscopic redshift catalog for all 89 sources robusted detected in the THRILS survey (Table \ref{tab:redshiftcatalog}).  This table also includes 13 additional sources with either continuum or emission-line detections but lacking a redshift measurement, as described in \S\ref{sec:redshifts}.  Included columns are the MPT ID from THRILS, RA and Dec coordinates of the source, a measured spectroscopic redshift ($z_{spec}$) from the THRILS data, a $z_{spec}$ quality grade, the total integration time on source, the target category (Primary, Filler, or Contaminant), photometry in 3 bandpasses (m$_H$, m$_{3.6}$, and m$_{4.5}$), and any prior photometric ($z_{phot}$) and spectroscopic ($z_{spec}$) redshift information available (as described in \S\ref{sec:catalogcreation}). Sources that have \hst-only derived $z_{phot}$ measurements, as they lie outside the CEERS NIRCam coverage \citep{finkelstein25}, are marked with an asterisk (*). Table footnotes denote the reference catalogs or publications from which the previous $z_{phot}$ and $z_{spec}$ measurements are taken. \\

\noindent Separately, we provide a table of the 81 non-detections in THRILS (Table \ref{tab:nondetections}) as described in \S\ref{sec:nondetections}. The format of this table is the same as Table \ref{tab:redshiftcatalog}; however we omit the redshift measurement and grade. \\

\noindent Machine-readable tables of both the spectroscopic redshift and non-detection catalogs are provided with the publication of this paper. 

\include{RedshiftCatalog}

\include{NonDetections}

\bibliography{papers}{}
\bibliographystyle{aasjournalv7}

\allauthors

~\vspace{1cm}\\
$^*$ NASA Postdoctoral Fellow\\
$^\dagger$ Giacconi Postdoctoral Fellow\\
$^\ddagger$ NSF Graduate Research Fellow\\
$^\S$ Dunlap Fellow\\
$^\P$ NSF Postdoctoral Research Fellow \\


\end{document}

%% file: commands.tex
\definecolor{notes}{HTML}{C70039}
\definecolor{emphasis_box}{HTML}{D1E1EC}
\definecolor{softgreen}{HTML}{468465}

\newcommand{\hb}{\hbox{\sc H$\beta$}}       
\newcommand{\oiii}{\hbox{\sc [O\,iii]}}     
\newcommand{\ha}{\hbox{\sc H$\alpha$}}      
\newcommand{\hd}{\hbox{\sc H$\delta$}} 
\newcommand{\nii}{\hbox{[N\,{\sc ii}]}}     
\newcommand{\sii}{\hbox{[S\,{\sc ii}]}}     
\newcommand{\siii}{\hbox{[S\,{\sc iii}]}}   
\newcommand{\hei}{He{\sc i}}

\newcommand{\paE}{\hbox{Pa$\epsilon$}}

\newcommand{\lam}{$\lambda$}
\newcommand{\unit}[1]{\ensuremath{\mathrm{\,#1}}\xspace}

\newcommand{\e}{\unit{e^{-}}}

\newcommand{\hst}{{\it HST}}

\newcommand{\jwst}{{\it JWST}}

\newcommand{\spitzer}{{\it Spitzer}}



%% file: RedshiftCatalog.tex

\begin{longrotatetable}
\begin{deluxetable*}{c|cc|cccc|ccc|cc}
\tablecolumns{12}
\tablecaption{The THRILS Survey Redshift Catalog}
\tablehead{
    \colhead{\textbf{THRILS}} & \multicolumn{2}{c}{\textbf{Coordinates}} & \multicolumn{4}{c}{\textbf{THRILS}} & \multicolumn{3}{c}{\textbf{Magnitudes}} & \multicolumn{2}{c}{\textbf{Prior Redshifts}}  \\ 
    \colhead{ID$_{MPT}$} & \colhead{RA} & \colhead{DEC} & \colhead{z$_{spec}$} & \colhead{Grade} & \colhead{ExpTime [s]} & \colhead{Category} & \colhead{H} & \colhead{3.60} & \colhead{4.50} & \colhead{z$_{phot}$} & \colhead{z$_{spec}$}} 
    \startdata
	102659 & 214.943146 & 52.94244 & 11.41 & 2 & 31862 & Primary & 28.4$\pm$0.1 & 27.95$\pm$0.04 & 27.68$\pm$0.04 & 11.6$^{10.0}_{10.0}$ & 11.42$^a$ \\
	117286 & 214.977181 & 52.926547 & 9.0307 & 1 & 15931 & Filler & 28.6$\pm$0.1 & 28.78$\pm$0.09 & 29.0$\pm$0.2 & 9.0$^{9.9}_{1.4}$ & -1 \\
	92641 & 214.823994 & 52.884186 & 8.93 & 1 & 15056 & Filler & 28.78$\pm$0.06 & 28.86$\pm$0.04 & 28.9$\pm$0.05 & 8.0$^{8.6}_{1.8}$ & -1 \\
	97442 & 214.838696 & 52.882226 & 8.929 & 3 & 30111 & Primary & 29.2$\pm$0.1 & 28.96$\pm$0.06 & 28.28$\pm$0.03 & 9.0$^{10.0}_{8.7}$ & -1 \\
	37717 & 214.93864 & 52.911749 & 8.7578 & 3 & 31862 & Primary & 26.48$\pm$0.02 & 26.76$\pm$0.01 & 26.24$\pm$0.009 & 9.0$^{9.0}_{8.8}$ & 8.768$^b$ \\
	19512 & 214.967531 & 52.932951 & 8.7146 & 3 & 31862 & Primary & 26.1$\pm$0.02 & 26.46$\pm$0.01 & 26.1$\pm$0.008 & 9.0$^{8.8}_{8.6}$ & 8.717$^b$ \\
	81287 & 214.995191 & 52.903856 & 8.6679 & 3 & 31862 & Primary & 28.19$\pm$0.09 & 28.58$\pm$0.07 & 27.84$\pm$0.05 & 10.5$^{10.0}_{9.0}$ & -1 \\
	115593 & 214.855926 & 52.846312 & 8.658 & 2 & 15056 & Filler & 29.4$\pm$0.1 & 29.7$\pm$0.09 & 29.08$\pm$0.06 & 8.0$^{9.3}_{8.1}$ & -1 \\
	107103 & 214.944763 & 52.931451 & 7.9229 & 3 & 31862 & Primary & 27.94$\pm$0.04 & 28.63$\pm$0.04 & 27.93$\pm$0.02 & 8.0$^{8.2}_{7.9}$ & 7.921$^b$ \\
	146988 & 214.992762 & 52.930886 & 7.9 & 3 & 15931 & Filler & 26.5$\pm$0.1 & -1 & -1 & 6.0$^{7.6}_{1.4}$* & -1 \\
	111728 & 214.911092 & 52.897315 & 7.4788 & 3 & 31862 & Primary & 27.97$\pm$0.04 & 28.25$\pm$0.03 & 27.4$\pm$0.02 & 7.0$^{7.4}_{7.3}$ & -1 \\
	117694 & 214.914025 & 52.88102 & 7.146 & 3 & 15056 & Primary & 28.75$\pm$0.09 & 28.5$\pm$0.04 & 27.81$\pm$0.03 & 7.0$^{7.4}_{7.2}$ & -1 \\
	44534 & 214.837993 & 52.863454 & 6.967 & 3 & 15056 & Primary & 27.49$\pm$0.05 & 26.73$\pm$0.02 & 26.62$\pm$0.01 & 7.0$^{7.4}_{6.9}$ & -1 \\
	96512 & 214.835418 & 52.882867 & 6.87 & 3 & 30111 & Filler & 28.36$\pm$0.09 & 28.81$\pm$0.09 & 29.2$\pm$0.1 & 7.0$^{7.4}_{6.8}$ & -1 \\
	44905 & 214.951073 & 52.93891 & 6.778 & 3 & 15931 & Filler & 28.69$\pm$0.06 & 28.14$\pm$0.02 & 28.74$\pm$0.05 & 7.0$^{6.8}_{6.7}$ & -1 \\
	46036 & 214.906056 & 52.893278 & 6.75 & 3 & 30111 & Primary & 27.9$\pm$0.1 & 26.89$\pm$0.02 & 27.57$\pm$0.03 & 7.0$^{6.7}_{6.7}$ & -1 \\
	47007 & 214.942703 & 52.905126 & 6.737 & 3 & 31862 & Primary & 28.6$\pm$0.1 & 27.2$\pm$0.02 & 28.13$\pm$0.04 & 7.0$^{6.8}_{6.7}$ & -1 \\
	16050 & 214.946731 & 52.900509 & 6.734 & 3 & 31862 & Primary & 26.1$\pm$0.02 & 25.48$\pm$0.009 & 25.79$\pm$0.01 & 7.0$^{6.9}_{6.8}$ & -1 \\
	11751 & 214.910857 & 52.852141 & 6.73 & 3 & 30111 & Primary & 24.39$\pm$0.01 & 25.87$\pm$0.03 & 26.28$\pm$0.06 & 6.0$^{6.3}_{6.2}$ & -1 \\
	114396 & 214.892867 & 52.86516 & 6.686 & 3 & 30111 & Primary & 29.0$\pm$0.07 & 28.47$\pm$0.03 & 29.32$\pm$0.06 & 7.0$^{6.7}_{6.6}$ & 6.7072$^b$ \\
	46403 & 214.892249 & 52.877403 & 6.68 & 3 & 30111 & Primary & 26.69$\pm$0.02 & 24.38$\pm$0.002 & 23.98$\pm$0.001 & 7.0$^{7.1}_{6.8}$ & 6.6832$^c$ \\
	43224 & 214.920965 & 52.940749 & 6.673 & 3 & 15931 & Filler & 27.41$\pm$0.04 & 26.59$\pm$0.01 & 26.99$\pm$0.01 & 7.0$^{6.7}_{1.4}$ & -1 \\
	44635 & 214.951621 & 52.943014 & 6.65 & 3 & 15931 & Filler & 27.36$\pm$0.05 & 26.83$\pm$0.02 & 27.38$\pm$0.04 & 7.0$^{6.7}_{6.6}$ & -1 \\
	25501 & 214.863029 & 52.889436 & 6.572 & 3 & 15056 & Primary & 26.19$\pm$0.02 & 25.29$\pm$0.008 & 25.92$\pm$0.009 & 7.0$^{6.6}_{6.6}$ & 6.579$^c$ \\
	44762 & 214.888951 & 52.896706 & 6.534 & 3 & 30111 & Primary & 27.32$\pm$0.04 & 26.81$\pm$0.01 & 27.23$\pm$0.02 & 7.0$^{6.6}_{6.5}$ & -1 \\
	48544 & 214.912338 & 52.858846 & 6.53 & 3 & 30111 & Primary & 28.34$\pm$0.07 & 27.44$\pm$0.02 & 28.28$\pm$0.05 & 7.0$^{6.7}_{6.6}$ & -1 \\
	23361 & 214.880968 & 52.891218 & 6.483 & 3 & 30111 & Primary & 26.33$\pm$0.02 & 25.52$\pm$0.007 & 26.16$\pm$0.01 & 7.0$^{6.6}_{6.6}$ & 6.4858$^c$ \\
	82535 & 214.996363 & 52.907148 & 6.459 & 2 & 15931 & Filler & 23.14$\pm$0.02 & 25.23$\pm$0.07 & 25.14$\pm$0.07 & 6.0$^{6.3}_{6.2}$ & -1 \\
	119163 & 214.949399 & 52.901025 & 6.21 & 2 & 15931 & Filler & 29.6$\pm$0.7 & 27.6$\pm$0.06 & 29.4$\pm$0.4 & 6.8$^{10.0}_{4.7}$ & -1 \\
	34328 & 214.991306 & 52.890046 & 6.183 & 3 & 31862 & Primary & 27.3$\pm$0.03 & 26.28$\pm$0.008 & 27.01$\pm$0.02 & 6.0$^{6.4}_{6.2}$ & 6.1872$^c$ \\
	140096 & 214.967647 & 52.938116 & 6.1385 & 2 & 15931 & Filler & 26.82$\pm$0.07 & -1 & 25.8$\pm$0.5 & 6.0$^{7.0}_{0.81}$* & -1 \\
	82330 & 214.910771 & 52.847172 & 6.011 & 3 & 15056 & Filler & 29.3$\pm$0.2 & 28.21$\pm$0.05 & 28.07$\pm$0.05 & 6.0$^{6.0}_{3.9}$ & -1 \\
	97420 & 214.855414 & 52.894118 & 5.993 & 3 & 30111 & Filler & 28.5$\pm$0.2 & 27.82$\pm$0.08 & 29.0$\pm$0.2 & 7.0$^{6.7}_{6.2}$ & -1 \\
	22417 & 214.89087 & 52.893244 & 5.98 & 3 & 30111 & Primary & 26.25$\pm$0.02 & 25.71$\pm$0.008 & 25.82$\pm$0.008 & 6.0$^{6.4}_{6.2}$ & 5.986$^c$ \\
	48907 & 214.975838 & 52.897791 & 5.94 & 3 & 15931 & Filler & 26.47$\pm$0.03 & 25.64$\pm$0.01 & 26.25$\pm$0.02 & 6.0$^{6.4}_{6.1}$ & -1 \\
	49055 & 214.903689 & 52.844912 & 5.805 & 3 & 30111 & Primary & 27.8$\pm$0.05 & 27.0$\pm$0.02 & 27.45$\pm$0.03 & 6.0$^{6.5}_{6.2}$ & 5.805$^b$ \\
	44025 & 214.823382 & 52.860373 & 5.751 & 3 & 30111 & Primary & 26.85$\pm$0.04 & 26.58$\pm$0.02 & 26.85$\pm$0.03 & 6.0$^{6.2}_{6.0}$ & 5.7558$^c$ \\
	49043 & 214.995872 & 52.910405 & 5.683 & 3 & 31862 & Primary & 27.41$\pm$0.03 & 26.79$\pm$0.01 & 27.06$\pm$0.02 & 6.0$^{5.9}_{5.6}$ & -1 \\
	78835 & 214.986859 & 52.891884 & 5.675 & 3 & 15931 & Filler & 29.4$\pm$0.3 & 27.64$\pm$0.03 & 27.89$\pm$0.05 & 6.0$^{6.3}_{4.5}$ & -1 \\
	87309 & 214.976767 & 52.90824 & 5.641 & 3 & 15931 & Filler & 29.0$\pm$-1.0 & 27.32$\pm$0.03 & 27.77$\pm$0.07 & 7.0$^{6.8}_{6.3}$ & -1 \\
	38124 & 214.95186 & 52.928259 & 5.482 & 3 & 31862 & Primary & 26.44$\pm$0.03 & 26.01$\pm$0.01 & 26.31$\pm$0.009 & 6.0$^{5.9}_{5.5}$ & 5.4805$^c$ \\
	101567 & 214.831734 & 52.867166 & 5.322 & 3 & 30111 & Primary & 28.59$\pm$0.09 & 28.54$\pm$0.06 & 27.85$\pm$0.03 & 7.0$^{7.4}_{7.0}$ & -1 \\
	29871 & 214.831188 & 52.886545 & 5.32 & 3 & 30111 & Primary & 25.72$\pm$0.02 & 24.95$\pm$0.007 & 24.89$\pm$0.006 & 5.0$^{5.5}_{5.4}$ & -1 \\
	45758 & 214.860947 & 52.864547 & 5.272 & 3 & 30111 & Primary & 28.1$\pm$0.1 & 27.19$\pm$0.03 & 27.22$\pm$0.04 & 5.0$^{5.3}_{1.5}$ & -1 \\
	40467 & 214.923949 & 52.948385 & 5.2715 & 3 & 31862 & Primary & 26.75$\pm$0.03 & 25.76$\pm$0.007 & 25.56$\pm$0.005 & 5.0$^{5.4}_{5.2}$ & 5.271$^c$ \\
	49400 & 215.000179 & 52.907742 & 5.12 & 3 & 31862 & Primary & 27.16$\pm$0.03 & 26.57$\pm$0.009 & 25.48$\pm$0.005 & 5.0$^{5.2}_{5.1}$ & -1 \\
	44774 & 214.828433 & 52.853442 & 5.101 & 3 & 30111 & Primary & 27.78$\pm$0.04 & 26.74$\pm$0.01 & 25.97$\pm$0.007 & 5.0$^{5.2}_{5.1}$ & -1 \\
	43139 & 214.926194 & 52.945551 & 5.087 & 3 & 31862 & Primary & 26.82$\pm$0.03 & 25.76$\pm$0.009 & 25.54$\pm$0.005 & 5.0$^{5.3}_{5.1}$ & -1 \\
	34913 & 214.981422 & 52.893632 & 5.077 & 3 & 15931 & Filler & 26.73$\pm$0.02 & 26.82$\pm$0.01 & 26.45$\pm$0.01 & 5.0$^{5.2}_{5.1}$ & -1 \\
	21489 & 214.926772 & 52.914039 & 5.071 & 3 & 31862 & Primary & 26.87$\pm$0.02 & 26.74$\pm$0.01 & 26.17$\pm$0.005 & 5.0$^{5.2}_{5.1}$ & -1 \\
	107963 & 214.839139 & 52.854132 & 4.99 & 3 & 30111 & Primary & 28.13$\pm$0.09 & 28.13$\pm$0.05 & 28.22$\pm$0.07 & 5.0$^{5.1}_{4.9}$ & -1 \\
	30374 & 214.915251 & 52.944979 & 4.894 & 3 & 31862 & Primary & 24.54$\pm$0.01 & 24.08$\pm$0.005 & 24.21$\pm$0.005 & 1.0$^{1.0}_{0.91}$ & -1 \\
	42648 & 214.913796 & 52.942971 & 4.89 & 3 & 31862 & Primary & 24.55$\pm$0.01 & 23.08$\pm$0.002 & 22.94$\pm$0.002 & 6.0$^{5.6}_{5.4}$ & -1 \\
	97936 & 214.917989 & 52.937245 & 4.889 & 3 & 31862 & Filler & 29.5$\pm$0.3 & 27.65$\pm$0.03 & 27.78$\pm$0.03 & 5.0$^{6.5}_{3.3}$ & 4.888$^b$ \\
	44527 & 214.942535 & 52.937728 & 4.887 & 3 & 10621 & Contaminant & 26.62$\pm$0.04 & 26.39$\pm$0.02 & 26.61$\pm$0.02 & 5.0$^{4.8}_{4.7}$ & -1 \\
	41237 & 214.917184 & 52.949369 & 4.88 & 3 & 31862 & Primary & 26.89$\pm$0.03 & 26.08$\pm$0.01 & 26.09$\pm$0.01 & 5.0$^{4.9}_{4.7}$ & -1 \\
	105834 & 214.8421 & 52.862287 & 4.81 & 1 & 30111 & Filler & -1 & 30.2$\pm$0.1 & 29.7$\pm$0.1 & 9.6$^{20.0}_{7.7}$ & -1 \\
	105963 & 214.889044 & 52.895062 & 4.809 & 3 & 15056 & Primary & 28.6$\pm$0.1 & 28.39$\pm$0.06 & 29.2$\pm$0.1 & 5.0$^{4.8}_{4.3}$ & -1 \\
	37813 & 214.898922 & 52.885389 & 4.808 & 3 & 30111 & Primary & 27.33$\pm$0.04 & 26.69$\pm$0.01 & 26.95$\pm$0.01 & 5.0$^{4.7}_{4.5}$ & 4.8082$^c$ \\
	100485 & 214.82091 & 52.861374 & 4.76 & 3 & 30111 & Primary & 28.6$\pm$0.2 & 27.33$\pm$0.04 & 27.62$\pm$0.05 & 5.0$^{4.8}_{4.3}$ & -1 \\
	17121 & 214.949419 & 52.907867 & 4.555 & 3 & 31862 & Primary & 25.1$\pm$0.01 & 24.14$\pm$0.003 & 24.22$\pm$0.004 & 5.0$^{4.8}_{4.7}$ & 4.561$^b$ \\
	43853 & 214.919841 & 52.931443 & 4.524 & 3 & 15931 & Filler & 27.1$\pm$0.07 & 26.77$\pm$0.01 & 27.58$\pm$0.02 & 5.0$^{4.8}_{4.6}$ & -1 \\
	47014 & 214.888082 & 52.866244 & 4.484 & 3 & 15056 & Filler & 28.08$\pm$0.05 & 27.39$\pm$0.02 & 27.53$\pm$0.02 & 4.0$^{4.6}_{4.1}$ & -1 \\
	22246 & 214.943956 & 52.929792 & 4.478 & 3 & 31862 & Primary & 25.12$\pm$0.02 & 23.61$\pm$0.003 & 23.77$\pm$0.003 & 4.0$^{4.6}_{4.3}$ & -1 \\
	43708 & 214.938903 & 52.946747 & 4.397 & 3 & 15931 & Filler & 28.0$\pm$0.1 & 27.41$\pm$0.03 & 27.46$\pm$0.04 & 5.0$^{4.7}_{4.2}$ & -1 \\
	105849 & 214.835706 & 52.857677 & 4.3952 & 2 & 15056 & Primary & 28.66$\pm$0.08 & 28.12$\pm$0.03 & 28.78$\pm$0.06 & 5.0$^{4.8}_{4.4}$ & -1 \\
	24975 & 214.940291 & 52.941319 & 4.39 & 3 & 31862 & Primary & 25.36$\pm$0.009 & 24.7$\pm$0.003 & 25.16$\pm$0.004 & 5.0$^{4.9}_{4.5}$* & -1 \\
	18709 & 214.858817 & 52.851472 & 4.355 & 3 & 30111 & Primary & 25.1$\pm$0.01 & 24.29$\pm$0.005 & 24.29$\pm$0.004 & 5.0$^{4.7}_{4.5}$ & 4.3618$^c$ \\
	20464 & 214.858837 & 52.860404 & 4.352 & 3 & 30111 & Filler & 25.35$\pm$0.02 & 23.26$\pm$0.002 & 23.12$\pm$0.002 & 2.0$^{4.2}_{2.5}$ & 4.3574$^c$ \\
	48965 & 214.973876 & 52.895808 & 4.114 & 3 & 31862 & Primary & 26.99$\pm$0.04 & 26.68$\pm$0.02 & 27.22$\pm$0.04 & 4.0$^{4.4}_{4.2}$ & -1 \\
	27259 & 214.918383 & 52.93789 & 3.912 & 3 & 31862 & Primary & 26.29$\pm$0.03 & 24.13$\pm$0.004 & 23.82$\pm$0.002 & 4.0$^{4.3}_{3.8}$ & -1 \\
	109756 & 214.887869 & 52.884643 & 3.7874 & 2 & 15056 & Filler & 30.7$\pm$0.4 & 30.19$\pm$0.08 & 31.4$\pm$0.3 & 5.0$^{6.7}_{0.82}$ & -1 \\
	24728 & 214.882084 & 52.89866 & 3.763 & 3 & 30111 & Primary & 24.9$\pm$0.01 & 23.86$\pm$0.003 & 23.71$\pm$0.002 & 4.0$^{4.2}_{3.9}$ & -1 \\
	118986 & 214.881081 & 52.852971 & 3.636 & 3 & 30111 & Filler & 28.28$\pm$0.09 & 28.14$\pm$0.05 & 28.26$\pm$0.06 & 4.0$^{4.4}_{3.7}$ & -1 \\
	13817 & 214.88459 & 52.844341 & 3.624 & 3 & 30111 & Primary & 24.84$\pm$0.007 & 24.46$\pm$0.005 & 24.27$\pm$0.003 & 4.0$^{4.3}_{4.2}$ & -1 \\
	46155 & 214.84549 & 52.848281 & 3.5179 & 3 & 30111 & Primary & 27.29$\pm$0.03 & 26.22$\pm$0.007 & 25.73$\pm$0.005 & 4.0$^{3.6}_{3.5}$ & 3.5186$^c$ \\
	38893 & 214.9361 & 52.930684 & 3.3759 & 3 & 15931 & Filler & 27.11$\pm$0.03 & 27.72$\pm$0.03 & 27.3$\pm$0.02 & 3.0$^{3.4}_{3.4}$ & -1 \\
	23103 & 214.940797 & 52.93229 & 3.3042 & 3 & 15931 & Filler & 26.62$\pm$0.02 & 27.35$\pm$0.02 & 27.06$\pm$0.01 & 3.0$^{3.4}_{3.4}$ & -1 \\
	16295 & 214.896148 & 52.865738 & 3.2387 & 3 & 15056 & Filler & 25.18$\pm$0.006 & 24.65$\pm$0.002 & 24.51$\pm$0.002 & 3.0$^{3.3}_{3.2}$ & -1 \\
	18945 & 214.917581 & 52.894662 & 3.218 & 3 & 15931 & Filler & 26.68$\pm$0.02 & 26.56$\pm$0.01 & 26.48$\pm$0.01 & 3.0$^{3.2}_{3.0}$ & -1 \\
	42514 & 214.853252 & 52.901552 & 3.025 & 3 & 30111 & Primary & 24.8$\pm$0.009 & 23.67$\pm$0.003 & 23.48$\pm$0.002 & 4.0$^{4.3}_{4.1}$ & -1 \\
	13658 & 214.974102 & 52.905943 & 2.6136 & 3 & 31862 & Primary & 22.14$\pm$0.002 & 20.71$\pm$0.0008 & 20.21$\pm$0.0007 & 2.0$^{2.4}_{2.3}$ & 2.6136$^c$ \\
	46649 & 214.857621 & 52.849534 & 1.918 & 2 & 10037 & Contaminant & 24.5$\pm$0.007 & 24.63$\pm$0.006 & 24.66$\pm$0.005 & 2.0$^{1.9}_{1.7}$ & -1 \\
	23870 & 214.950243 & 52.942729 & 1.279 & 1 & 10621 & Contaminant & 25.47$\pm$0.01 & 25.53$\pm$0.01 & 25.62$\pm$0.01 & 1.0$^{1.3}_{1.2}$ & -1 \\
	29328 & 214.834268 & 52.893785 & 0.7 & 2 & 30111 & Primary & 19.17$\pm$0.001 & 18.79$\pm$0.001 & 19.2$\pm$0.0007 & 0.7$^{0.73}_{0.67}$ & 0.7328$^d$ \\
	137586 & 215.003142 & 52.896297 & 0.0 & 1 & 5310 & Contaminant & 16.15$\pm$8e-06 & 16.93$\pm$0.0001 & 17.31$\pm$0.0002 & 0.01$^{0.02}_{0.01}$ & -1 \\
	156965 & 214.987478 & 52.89089 & 0.0 & 1 & 31862 & Primary & 19.06$\pm$0.0002 & 19.85$\pm$0.002 & 20.18$\pm$0.002 & 0.04$^{0.05}_{0.03}$* & -1 \\
	21240 & 214.947976 & 52.925768 & 0.0 & 1 & 31862 & Primary & 20.64$\pm$0.004 & 21.46$\pm$0.002 & 21.76$\pm$0.0007 & 0.01$^{0.04}_{0.01}$ & -1 \\
	17639 & 214.91385 & 52.884142 & 0.0 & 1 & 30111 & Primary & 21.24$\pm$0.002 & 22.14$\pm$0.0009 & 22.41$\pm$0.0006 & 6.0$^{6.2}_{6.1}$ & -1 \\
	136596 & 215.00334 & 52.895862 & -1.0 & 0 & 15931 & Filler & 22.05$\pm$0.003 & -1 & -1 & 5.0$^{4.6}_{4.5}$* & -1 \\
	27920 & 214.854892 & 52.896851 & -1.0 & 0 & 15056 & Filler & 25.48$\pm$0.009 & 25.27$\pm$0.006 & 25.39$\pm$0.006 & 4.0$^{4.2}_{4.0}$ & -1 \\
	46370 & 214.971028 & 52.933685 & -2.0 & 0 & 15931 & Filler & 25.16$\pm$0.01 & 25.05$\pm$0.008 & 25.19$\pm$0.02 & 1.0$^{1.3}_{1.1}$ & -1 \\
	14536 & 214.977299 & 52.913015 & -2.0 & 0 & 15931 & Filler & 27.02$\pm$0.05 & 27.04$\pm$0.03 & 27.1$\pm$0.05 & 0.0$^{2.7}_{0.19}$ & 2.2107$^e$ \\
	141538 & 214.983922 & 52.944899 & -2.0 & 0 & 15931 & Filler & 27.1$\pm$0.09 & 27.5$\pm$0.9 & 28.0$\pm$1.0 & 3.0$^{3.4}_{1.1}$* & -1 \\
	38368 & 214.96954 & 52.944872 & -2.0 & 0 & 15931 & Filler & 27.62$\pm$0.03 & 28.81$\pm$0.04 & 28.76$\pm$0.04 & 0.01$^{0.07}_{0.01}$ & -1 \\
	118431 & 214.918319 & 52.880803 & -2.0 & 0 & 15056 & Filler & 28.0$\pm$0.1 & 28.12$\pm$0.09 & 28.2$\pm$0.1 & 5.0$^{5.2}_{3.5}$ & -1 \\
	116521 & 214.888015 & 52.866342 & -2.0 & 0 & 10037 & Filler & 29.4$\pm$0.2 & 29.7$\pm$0.2 & 32.0$\pm$1.0 & 0.0$^{6.8}_{0.43}$ & -1 \\
	114003 & 214.887211 & 52.862043 & -2.0 & 0 & 30111 & Primary & 28.7$\pm$0.1 & 28.6$\pm$0.07 & 28.22$\pm$0.04 & 11.4$^{10.0}_{10.0}$ & 2.8884$^c$ \\
	9528 & 215.003141 & 52.896296 & -3.0 & 0 & 5310 & Contaminant & 16.0$\pm$-1.0 & 17.0$\pm$-1.0 & 18.0$\pm$-1.0 & 0.5$^{0.48}_{0.45}$* & -1 \\
	21981 & 214.9431 & 52.925335 & -3.0 & 0 & 15931 & Filler & 21.83$\pm$0.003 & 21.45$\pm$0.002 & 21.68$\pm$0.001 & 1.0$^{1.2}_{1.1}$ & -1 \\
	20223 & 214.912872 & 52.897786 & -3.0 & 0 & 15931 & Filler & 25.5$\pm$0.02 & 25.42$\pm$0.01 & 25.43$\pm$0.01 & 2.0$^{1.6}_{1.4}$ & -1 \\
	48447 & 214.916502 & 52.862885 & -3.0 & 0 & 30111 & Filler & 26.7$\pm$0.05 & 26.1$\pm$0.02 & 25.95$\pm$0.02 & 3.0$^{3.0}_{1.8}$ & -1 \\
    \enddata
\tablecomments{Full redshift catalog for the 89 sources with spectroscopic redshift measurements in the THRILS data, as well as the 13 additional sources with detectable lines ($z_{spec} = -2 $) or detected continuum ($z_{spec} = -3 $) but no discernible redshift. We note that for all $z_{spec}$ from the THRILS program, the redshift uncertainty is 0.0003. Sources with a $z_{spec} = 0$ are likely stars or brown dwarfs.  
Description of Columns: 1) ID in the THRILS MPT, 2-3) Coordinates, 4-5) Measured $z_{spec}$ in THRILS and grade for this redshift as described in \S\ref{sec:redshifts}, 6-7) Exposure time and target category for the source as described in \S\ref{sec:targetselection}, 8-10) Magnitude of the source in 3 representative bands as described in \S\ref{sec:catalogcreation}, 11-12) Prior measurements of $z_{phot}$ and $z_{spec}$ from ancillary data as described in \S\ref{sec:catalogcreation}.  \\
~\vspace{-2.5mm}\\
{\footnotesize * Sources with \hst\ only derived $z_{phot}$ measurements as they do not have \jwst/NIRCam coverage. \\
$^a$ \jwst\ DDT 2750 \citep{arrabalharo23}. \\
$^b$ CEERS (Arrabal Haro et al., in prep).\\ 
$^c$ DJA \citep{degraaff24, heintz24}. \\
$^d$ CANDELS \citep{stefanon17}. \\
$^e$ 3DHST \citep{brammer12} }} \label{tab:redshiftcatalog}
\end{deluxetable*} 
\end{longrotatetable}

%% file: NonDetections.tex
\startlongtable 
\begin{deluxetable*}{c|cc|cc|ccc|cc}\label{tab:nondetections}
\centerwidetable
\tablecolumns{10}
\tablecaption{Non-Detections from the THRILS Survey}
\tablehead{
    \colhead{\textbf{THRILS}} & \multicolumn{2}{c}{\textbf{Coordinates}} & \multicolumn{2}{c}{\textbf{THRILS}} & \multicolumn{3}{c}{\textbf{Magnitudes}} & \multicolumn{2}{c}{\textbf{Prior Redshifts}}  \\ 
    \colhead{ID$_{MPT}$} & \colhead{RA} & \colhead{DEC} & \colhead{ExpTime [s]} & \colhead{Category} & \colhead{H} & \colhead{3.60} & \colhead{4.50} & \colhead{z$_{phot}$} & \colhead{z$_{spec}$}} 
    \startdata
		127239 & 214.85961 & 52.858029 & 15056 & Filler & -1 & -1 & -1 & -1 & 6.0611$^a$ \\
	136697 & 215.003204 & 52.896604 & 15931 & Filler & 23.5$\pm$0.003 & -1 & -1 & 5.0$^{5.8}_{5.4}$* & -1 \\
	139799 & 214.948036 & 52.926073 & 10621 & Contaminant & 25.7$\pm$0.07 & -1 & -1 & 2.0$^{8.0}_{1.4}$* & -1 \\
	160204 & 214.841267 & 52.891728 & 15056 & Filler & 27.17$\pm$0.07 & -1 & -1 & 5.0$^{5.4}_{2.5}$* & -1 \\
	139366 & 214.9154 & 52.904181 & 15931 & Filler & 27.5$\pm$0.1 & -1 & -1 & 10.5$^{10.0}_{3.2}$* & -1 \\
	160727 & 214.913099 & 52.941113 & 15931 & Filler & 27.7$\pm$0.1 & -1 & -1 & 3.0$^{10.0}_{2.2}$* & -1 \\
	127087 & 214.949399 & 52.908 & 21241 & Contaminant & 29.0$\pm$-1.0 & -1 & -1 & 5.0$^{-1.0}_{-1.0}$* & -1 \\
	143034 & 214.921523 & 52.895192 & 31862 & Filler & 27.6$\pm$0.1 & 28.0$\pm$3.0 & -1 & 10.6$^{10.0}_{10.0}$* & -1 \\
	113779 & 214.86049 & 52.855005 & 30111 & Filler & 30.4$\pm$0.6 & 29.3$\pm$0.2 & -1 & 17.4$^{20.0}_{6.4}$ & -1 \\
	105741 & 214.88741 & 52.894714 & 10037 & Filler & 30.3$\pm$0.3 & 29.8$\pm$0.1 & -1 & 6.7$^{20.0}_{2.8}$ & -1 \\
	115794 & 214.9694 & 52.926311 & 10621 & Contaminant & 31.0$\pm$1.0 & 30.2$\pm$0.2 & -1 & 17.0$^{20.0}_{5.5}$ & -1 \\
	104831 & 214.831237 & 52.857013 & 10037 & Contaminant & 30.5$\pm$0.4 & 30.6$\pm$0.3 & -1 & 0.6$^{10.0}_{0.85}$ & -1 \\
	110862 & 214.947994 & 52.925387 & 10621 & Contaminant & 28.76$\pm$0.06 & 31.0$\pm$0.2 & -1 & 7.0$^{8.0}_{6.9}$ & -1 \\
	18658 & 214.857584 & 52.849377 & 5019 & Contaminant & 22.38$\pm$0.002 & 22.38$\pm$0.002 & 22.37$\pm$0.001 & 1.0$^{1.7}_{1.1}$ & 0.9137$^b$ \\
	160549 & 214.840565 & 52.89058 & 15056 & Filler & 27.1$\pm$0.1 & 23.86$\pm$0.07 & 23.69$\pm$0.06 & 5.0$^{4.8}_{2.8}$* & -1 \\
	40281 & 214.848324 & 52.89162 & 15056 & Filler & 28.6$\pm$0.2 & 33.0$\pm$9.0 & 25.0$\pm$-1.0 & 5.0$^{5.2}_{5.1}$ & -1 \\
	158240 & 214.83878 & 52.89779 & 15056 & Filler & 27.32$\pm$0.03 & 25.3$\pm$0.3 & 25.7$\pm$0.3 & 12.5$^{10.0}_{10.0}$* & -1 \\
	112097 & 214.913199 & 52.897737 & 15931 & Filler & 32.3$\pm$0.5 & 32.1$\pm$0.2 & 25.7$\pm$0.4 & 17.1$^{20.0}_{3.7}$ & -1 \\
	10801 & 214.998121 & 52.909333 & 5310 & Contaminant & 26.49$\pm$0.02 & 26.32$\pm$0.01 & 26.28$\pm$0.01 & 2.0$^{2.6}_{2.0}$ & -1 \\
	38054 & 214.933701 & 52.914233 & 15931 & Filler & 26.6$\pm$0.02 & 26.38$\pm$0.008 & 26.33$\pm$0.008 & 3.0$^{3.0}_{2.7}$ & -1 \\
	45468 & 214.926827 & 52.914352 & 10621 & Contaminant & 26.24$\pm$0.03 & 26.2$\pm$0.01 & 26.46$\pm$0.01 & 0.7$^{0.76}_{0.55}$ & -1 \\
	44514 & 214.942513 & 52.937886 & 15931 & Filler & 26.68$\pm$0.04 & 26.52$\pm$0.02 & 26.56$\pm$0.02 & 5.0$^{5.8}_{4.8}$ & -1 \\
	41708 & 214.839797 & 52.901349 & 30111 & Primary & 27.4$\pm$0.1 & 26.66$\pm$0.04 & 26.75$\pm$0.05 & 5.0$^{5.4}_{1.3}$ & -1 \\
	44772 & 214.93045 & 52.925902 & 15931 & Filler & 26.24$\pm$0.03 & 26.58$\pm$0.02 & 26.85$\pm$0.02 & 1.0$^{1.1}_{0.94}$ & -1 \\
	141186 & 214.920742 & 52.901108 & 15931 & Filler & 27.5$\pm$0.1 & 26.0$\pm$2.0 & 27.0$\pm$7.0 & 12.1$^{10.0}_{2.7}$* & -1 \\
	46784 & 214.969498 & 52.927092 & 10621 & Filler & 27.07$\pm$0.04 & 27.64$\pm$0.04 & 27.56$\pm$0.04 & 1.0$^{1.8}_{1.3}$ & -1 \\
	21020 & 214.888957 & 52.884764 & 10037 & Filler & 27.0$\pm$0.2 & 26.98$\pm$0.04 & 27.61$\pm$0.06 & 0.4$^{0.46}_{0.25}$ & -1 \\
	31210 & 214.915047 & 52.948342 & 10621 & Contaminant & 27.01$\pm$0.03 & 27.52$\pm$0.03 & 27.7$\pm$0.04 & 2.0$^{2.1}_{1.7}$ & -1 \\
	112432 & 214.965906 & 52.933838 & 15931 & Filler & 27.55$\pm$0.06 & 27.67$\pm$0.04 & 27.79$\pm$0.06 & 0.0$^{1.9}_{0.91}$ & -1 \\
	112876 & 214.900136 & 52.886169 & 30111 & Filler & 27.77$\pm$0.06 & 27.74$\pm$0.03 & 27.85$\pm$0.04 & 1.0$^{1.7}_{0.97}$ & -1 \\
	104864 & 214.831294 & 52.856964 & 15056 & Filler & -1 & 28.6$\pm$0.2 & 27.9$\pm$0.1 & 15.7$^{20.0}_{9.3}$ & -1 \\
	104854 & 214.841842 & 52.864873 & 15056 & Filler & 27.54$\pm$0.04 & 27.78$\pm$0.03 & 27.91$\pm$0.04 & 2.0$^{2.4}_{1.8}$ & -1 \\
	117548 & 214.944021 & 52.903844 & 15931 & Filler & 28.4$\pm$0.2 & 28.52$\pm$0.09 & 27.97$\pm$0.09 & 6.0$^{7.4}_{4.6}$ & -1 \\
	84860 & 214.902185 & 52.848443 & 30111 & Filler & 26.64$\pm$0.05 & 28.0$\pm$0.1 & 28.0$\pm$0.1 & 6.0$^{6.4}_{6.0}$ & -1 \\
	101330 & 214.932228 & 52.938464 & 15931 & Filler & 27.8$\pm$0.1 & 27.91$\pm$0.08 & 28.02$\pm$0.06 & 6.0$^{6.2}_{5.1}$ & -1 \\
	94028 & 214.82841 & 52.883933 & 15056 & Filler & 28.2$\pm$0.1 & 28.07$\pm$0.06 & 28.04$\pm$0.06 & 6.0$^{6.1}_{1.4}$ & -1 \\
	86451 & 214.972369 & 52.902641 & 15931 & Filler & 28.7$\pm$0.2 & 28.8$\pm$0.08 & 28.15$\pm$0.07 & 8.0$^{8.6}_{5.8}$ & -1 \\
	82490 & 214.998193 & 52.909699 & 15931 & Filler & 28.21$\pm$0.07 & 27.9$\pm$0.03 & 28.16$\pm$0.05 & 1.0$^{1.5}_{1.0}$ & -1 \\
	100820 & 214.855425 & 52.885063 & 30111 & Primary & 28.0$\pm$0.1 & 28.8$\pm$0.1 & 28.25$\pm$0.07 & 9.8$^{10.0}_{9.0}$ & -1 \\
	104352 & 214.950295 & 52.942885 & 15931 & Filler & 29.0$\pm$0.2 & 29.3$\pm$0.1 & 28.34$\pm$0.07 & 8.0$^{7.6}_{7.2}$ & -1 \\
	123654 & 214.973961 & 52.905454 & 31862 & Filler & 31.0$\pm$1.0 & 30.4$\pm$0.4 & 28.35$\pm$0.09 & 14.7$^{20.0}_{10.0}$ & -1 \\
	119258 & 214.881119 & 52.853086 & 20074 & Contaminant & 27.48$\pm$0.06 & 28.31$\pm$0.07 & 28.38$\pm$0.08 & 0.0$^{3.2}_{0.25}$ & -1 \\
	115974 & 214.969318 & 52.925426 & 10621 & Contaminant & 28.0$\pm$0.1 & 28.7$\pm$0.2 & 28.4$\pm$0.2 & 6.0$^{8.0}_{2.6}$ & -1 \\
	86348 & 214.916438 & 52.862712 & 20074 & Contaminant & 28.2$\pm$0.2 & 28.3$\pm$0.1 & 28.4$\pm$0.2 & 0.2$^{10.0}_{1.2}$ & -1 \\
	87467 & 214.886685 & 52.844783 & 15056 & Filler & 28.28$\pm$0.09 & 28.13$\pm$0.05 & 28.51$\pm$0.07 & 5.0$^{4.9}_{0.88}$ & -1 \\
	116137 & 214.857639 & 52.849729 & 15056 & Filler & 27.0$\pm$-1.0 & 27.81$\pm$0.08 & 28.6$\pm$0.2 & 4.8$^{10.0}_{1.2}$ & -1 \\
	81605 & 215.001818 & 52.909548 & 15931 & Filler & 28.54$\pm$0.07 & 28.53$\pm$0.04 & 28.69$\pm$0.06 & 2.0$^{2.7}_{1.9}$ & -1 \\
	123181 & 214.96947 & 52.926907 & 15931 & Filler & 28.0$\pm$-1.0 & -1 & 28.8$\pm$0.1 & 9.8$^{10.0}_{8.8}$ & -1 \\
	123813 & 214.976084 & 52.900459 & 15931 & Filler & -1 & -1 & 28.8$\pm$0.08 & 9.8$^{10.0}_{9.7}$ & -1 \\
	113467 & 214.904735 & 52.887273 & 15056 & Filler & 28.8$\pm$0.2 & 28.56$\pm$0.06 & 28.8$\pm$0.08 & 0.0$^{2.1}_{0.76}$ & -1 \\
	114210 & 214.945117 & 52.913788 & 15931 & Filler & 27.9$\pm$0.1 & 28.09$\pm$0.09 & 28.8$\pm$0.2 & 7.0$^{8.2}_{1.8}$ & -1 \\
	84761 & 214.902143 & 52.848134 & 10037 & Contaminant & 28.8$\pm$0.2 & 28.6$\pm$0.1 & 28.8$\pm$0.2 & 6.0$^{6.3}_{1.6}$ & -1 \\
	119213 & 214.94987 & 52.901436 & 15931 & Filler & 29.0$\pm$0.1 & 29.01$\pm$0.06 & 28.85$\pm$0.08 & 8.0$^{8.6}_{1.7}$ & -1 \\
	101212 & 214.932314 & 52.938602 & 10621 & Filler & 28.3$\pm$0.1 & 28.62$\pm$0.09 & 28.9$\pm$0.1 & 0.0$^{3.1}_{0.31}$ & -1 \\
	86560 & 214.971243 & 52.902124 & 15931 & Filler & 28.0$\pm$-1.0 & 27.6$\pm$0.1 & 28.9$\pm$0.5 & 19.3$^{20.0}_{4.2}$ & -1 \\
	114558 & 214.97158 & 52.931663 & 15931 & Filler & 29.2$\pm$0.4 & 28.2$\pm$0.1 & 28.9$\pm$0.3 & 7.0$^{6.9}_{5.6}$ & -1 \\
	97545 & 214.855401 & 52.893852 & 10037 & Contaminant & 29.4$\pm$0.2 & 28.61$\pm$0.06 & 28.95$\pm$0.07 & 5.0$^{5.8}_{3.7}$ & -1 \\
	150502 & 214.984105 & 52.912764 & 15931 & Filler & 27.42$\pm$0.07 & -1 & 29.0$\pm$7.0 & 12.1$^{10.0}_{10.0}$* & -1 \\
	105717 & 214.887338 & 52.894826 & 15056 & Filler & 30.4$\pm$0.5 & 29.6$\pm$0.2 & 29.02$\pm$0.08 & 15.3$^{20.0}_{3.3}$ & -1 \\
	123426 & 214.949958 & 52.899793 & 15931 & Filler & 33.0$\pm$5.0 & 30.7$\pm$0.2 & 29.04$\pm$0.09 & 15.3$^{20.0}_{7.5}$ & -1 \\
	123055 & 214.860476 & 52.854767 & 20074 & Contaminant & 31.0$\pm$1.0 & 29.1$\pm$0.1 & 29.1$\pm$0.1 & 4.0$^{20.0}_{3.8}$ & -1 \\
	97782 & 214.911942 & 52.933363 & 15931 & Filler & 28.3$\pm$0.3 & 28.6$\pm$0.2 & 29.1$\pm$0.3 & 9.4$^{20.0}_{4.2}$ & -1 \\
	123620 & 214.962746 & 52.897695 & 15931 & Filler & -1 & 30.7$\pm$0.3 & 29.1$\pm$0.1 & 15.8$^{20.0}_{7.8}$ & -1 \\
	82395 & 214.910708 & 52.847336 & 10037 & Filler & 29.1$\pm$0.1 & 29.17$\pm$0.09 & 29.3$\pm$0.1 & 1.0$^{5.3}_{1.2}$ & -1 \\
	107965 & 214.834643 & 52.850082 & 15056 & Filler & -1 & 29.1$\pm$0.2 & 29.3$\pm$0.3 & 16.9$^{20.0}_{4.4}$ & -1 \\
	123630 & 214.973897 & 52.905182 & 10621 & Contaminant & 29.0$\pm$-1.0 & -1 & 29.4$\pm$0.1 & 9.8$^{10.0}_{8.9}$ & -1 \\
	116741 & 214.888132 & 52.8666 & 5019 & Contaminant & 29.9$\pm$0.2 & 29.7$\pm$0.1 & 29.43$\pm$0.07 & 6.0$^{6.3}_{3.0}$ & -1 \\
	125652 & 214.888834 & 52.884585 & 15056 & Filler & -1 & 30.5$\pm$0.3 & 29.4$\pm$0.1 & 17.6$^{20.0}_{8.7}$ & -1 \\
	117080 & 214.944081 & 52.904108 & 5310 & Contaminant & 30.1$\pm$0.3 & 29.38$\pm$0.06 & 29.5$\pm$0.1 & 4.0$^{6.8}_{3.8}$ & -1 \\
	88081 & 214.977307 & 52.913208 & 10621 & Contaminant & 30.0$\pm$0.3 & 29.6$\pm$0.1 & 29.6$\pm$0.2 & 6.0$^{6.7}_{2.1}$ & -1 \\
	84338 & 214.968669 & 52.894157 & 15931 & Filler & 31.0$\pm$-1.0 & 28.81$\pm$0.08 & 29.7$\pm$0.3 & 13.7$^{10.0}_{10.0}$ & -1 \\
	105649 & 214.887413 & 52.895097 & 5019 & Contaminant & 29.1$\pm$0.1 & 29.9$\pm$0.1 & 29.8$\pm$0.1 & 1.0$^{4.2}_{1.1}$ & -1 \\
	93662 & 214.888977 & 52.895245 & 10037 & Filler & 29.0$\pm$0.1 & 29.6$\pm$0.1 & 29.8$\pm$0.1 & 0.5$^{0.82}_{0.28}$ & -1 \\
	106298 & 214.845606 & 52.863064 & 15056 & Filler & 31.0$\pm$2.0 & 29.1$\pm$0.1 & 30.0$\pm$0.4 & 15.5$^{20.0}_{10.0}$ & -1 \\
	105874 & 214.891941 & 52.897584 & 15056 & Filler & 31.0$\pm$2.0 & 29.0$\pm$0.1 & 30.0$\pm$0.3 & 4.0$^{4.5}_{0.67}$ & -1 \\
	102778 & 214.880983 & 52.898025 & 15056 & Filler & 29.4$\pm$0.3 & 28.9$\pm$0.1 & 30.2$\pm$0.3 & 1.0$^{4.6}_{0.55}$ & -1 \\
	106725 & 214.841309 & 52.858729 & 15056 & Filler & 29.0$\pm$0.3 & 28.4$\pm$0.1 & 30.2$\pm$0.6 & 7.0$^{7.7}_{1.9}$ & -1 \\
	106752 & 214.841367 & 52.858767 & 10037 & Filler & 29.1$\pm$0.2 & 30.6$\pm$0.5 & 30.2$\pm$0.4 & 7.0$^{9.1}_{6.1}$ & -1 \\
	87731 & 214.968843 & 52.944907 & 15931 & Filler & 30.1$\pm$0.4 & 29.7$\pm$0.1 & 30.6$\pm$0.4 & 5.0$^{5.4}_{1.3}$ & -1 \\
	105221 & 214.936107 & 52.930431 & 5310 & Contaminant & 30.1$\pm$0.2 & 30.7$\pm$0.1 & 31.4$\pm$0.2 & 0.0$^{3.1}_{0.13}$ & -1 \\
	120077 & 214.89006 & 52.85531 & 15056 & Filler & 27.6$\pm$0.1 & 27.92$\pm$0.09 & 31.0$\pm$3.0 & 4.0$^{4.5}_{0.64}$ & -1 \\
	106646 & 214.838367 & 52.856751 & 15056 & Filler & -1 & 28.41$\pm$0.08 & 32.0$\pm$3.0 & 19.2$^{20.0}_{10.0}$ & -1 \\
    \enddata
\tablecomments{Table of the 81 non-detections in the THRILS data sorted in order of brightness in three filters (in descending wavelength order). Sources that are not detected or do not have data in a certain filter are assigned a value of -1 for that magnitude. 
    Description of Columns: 1) ID in the THRILS MPT, 2-3) Coordinates, 4--5) Exposure time and target category for the source as described in \S\ref{sec:targetselection}, 6--8) Magnitude of the source in 3 representative bands as described in \S\ref{sec:catalogcreation}, 9--10) Prior measurements of $z_{phot}$ and $z_{spec}$ from ancillary data as described in \S\ref{sec:catalogcreation}. \\
~\vspace{-2.5mm}\\
{\footnotesize * Sources with \hst\ only derived $z_{phot}$ measurements as they do not have \jwst/NIRCam coverage. \\
$^a$ DJA \citep{degraaff24, heintz24}. \\
$^b$ CANDELS DEEP2 \citep{stefanon17}. } }
    
\end{deluxetable*} 